\documentclass[11pt,letterpaper]{article}
\usepackage{fullpage}
\usepackage[utf8]{inputenc}

\usepackage{amsmath}
\usepackage{amsfonts}
\usepackage{amssymb}
\usepackage{amsthm}
\usepackage{bm}
\usepackage{natbib}
\usepackage{url}
\usepackage[colorlinks,linkcolor=blue,citecolor=blue,urlcolor=blue]{hyperref}
\usepackage{mathtools}
\usepackage{authblk}
\usepackage{graphicx}
\usepackage[usenames,dvipsnames]{xcolor}
\usepackage{verbatim}
\usepackage{color}
\usepackage{soul}
\usepackage{setspace}
\usepackage{algorithm}
\usepackage{algpseudocode}
\usepackage{subcaption}
\usepackage{booktabs}

\newtheorem{definition}{Definition}

\newcommand{\cb}{\bm{c}}
\newcommand{\Cb}{\bm{C}}

\newcommand{\Db}{\bm{D}}

\newcommand{\fb}{\bm{f}}

\newcommand{\Ib}{\bm{I}}

\newcommand{\mb}{\bm{m}}

\newcommand{\Pb}{\bm{P}}

\newcommand{\ub}{\bm{u}}
\newcommand{\Ub}{\bm{U}}

\newcommand{\Vb}{\bm{V}}

\newcommand{\Wb}{\bm{W}}
\newcommand{\xb}{\bm{x}}

\newcommand{\zb}{\bm{z}}
\newcommand{\Zb}{\bm{Z}}

\newcommand{\deltab}{\bm{\delta}}
\newcommand{\epsilonb}{\bm{\epsilon}}
\newcommand{\varepsilonb}{\bm{\varepsilon}}

\newcommand{\etab}{\bm{\eta}}
\newcommand{\thetab}{\bm{\theta}}

\newcommand{\lambdab}{\bm{\lambda}}
\newcommand{\mub}{\bm{\mu}}

\newcommand{\phib}{\bm{\phi}}

\newcommand{\Lambdab}{\bm{\Lambda}}
\newcommand{\Xib}{\bm{\Xi}}

\newcommand{\Sigmab}{\bm{\Sigma}}

\newcommand{\Psib}{\bm{\Psi}}
\newcommand{\Omegab}{\bm{\Omega}}

\newcommand{\E}{\mbox{E}}
\newcommand{\Var}{\mbox{Var}}
\newcommand{\cov}{\mbox{cov}}
\newcommand{\diag}{\mbox{diag}}

\newcommand{\R}{\mathbb{R}}

\def\NPN{\textup{NPN}}

\def\N{\textup{N}}

\def\IG{\textup{IG}}

\def\giG{\textup{giG}}

\begin{document}
\title{Bayesian segmented Gaussian copula factor model for single-cell sequencing data}
\author[1]{Junsouk Choi \thanks{junsouk@umich.edu}}
\author[2]{Hee Cheol Chung}
\author[1]{Irina Gaynanova}
\author[3]{Yang Ni}
\affil[1]{\small Department of Biostatistics, University of Michigan, Ann Arbor, Michigan, U.S.A.}
\affil[2]{\small Department of Mathematics and Statistics, University of North Carolina at Charlotte, Charlotte, NC 28223, USA}
\affil[3]{\small Department of Statistics, Texas A\&M University, College Station, TX 77843, USA}
\date{}
\setstretch{1.5}
\maketitle
\begin{abstract}
Single-cell sequencing technologies have significantly advanced molecular and cellular biology, offering unprecedented insights into cellular heterogeneity by allowing for the measurement of gene expression at an individual cell level. However, the analysis of such data is challenged by the prevalence of low counts due to dropout events and the skewed nature of the data distribution, which conventional Gaussian factor models struggle to handle effectively. To address these challenges, we propose a novel Bayesian segmented Gaussian copula model to explicitly account for inflation of zero and near-zero counts, and to address the high skewness in the data. By employing a Dirichlet-Laplace prior for each column of the factor loadings matrix, we shrink the loadings of unnecessary factors towards zero, which leads to a simple approach to automatically determine the number of latent factors, and resolve the identifiability issue inherent in factor models due to the rotational invariance of the factor loadings matrix.  Through simulation studies, we demonstrate the superior performance of our method over existing approaches in conducting factor analysis on data exhibiting the characteristics of single-cell data, such as excessive low counts and high skewness. Furthermore, we apply the proposed method to a real single-cell RNA-sequencing dataset from a lymphoblastoid cell line, successfully identifying biologically meaningful latent factors and detecting previously uncharacterized cell subtypes.
\end{abstract}

\clearpage
\doublespacing

\section{Introduction}
\label{sec:intro}

Factor analysis or latent factor model is a powerful tool for multivariate data exploration, routinely applied in fields such as social science, economics, and genomics. 
It reduces dimension by representing observed variables in terms of a linear combination of a smaller number of latent factors, simplifying the interpretation of complex datasets. In this paper, we focus on single-cell sequencing data, with a motivating dataset from the 10x Genomics single-cell RNA-sequencing (scRNA-seq). Our goal is to develop a novel factor analysis approach for single-cell sequencing data, aimed at uncovering the underlying dependencies among sequenced genes that reflect coordinated biological processes, while simultaneously reducing the dimensionality of the data to facilitate downstream analyses and data visualization.

\begin{figure}
\centering
\includegraphics[width=0.9\linewidth]{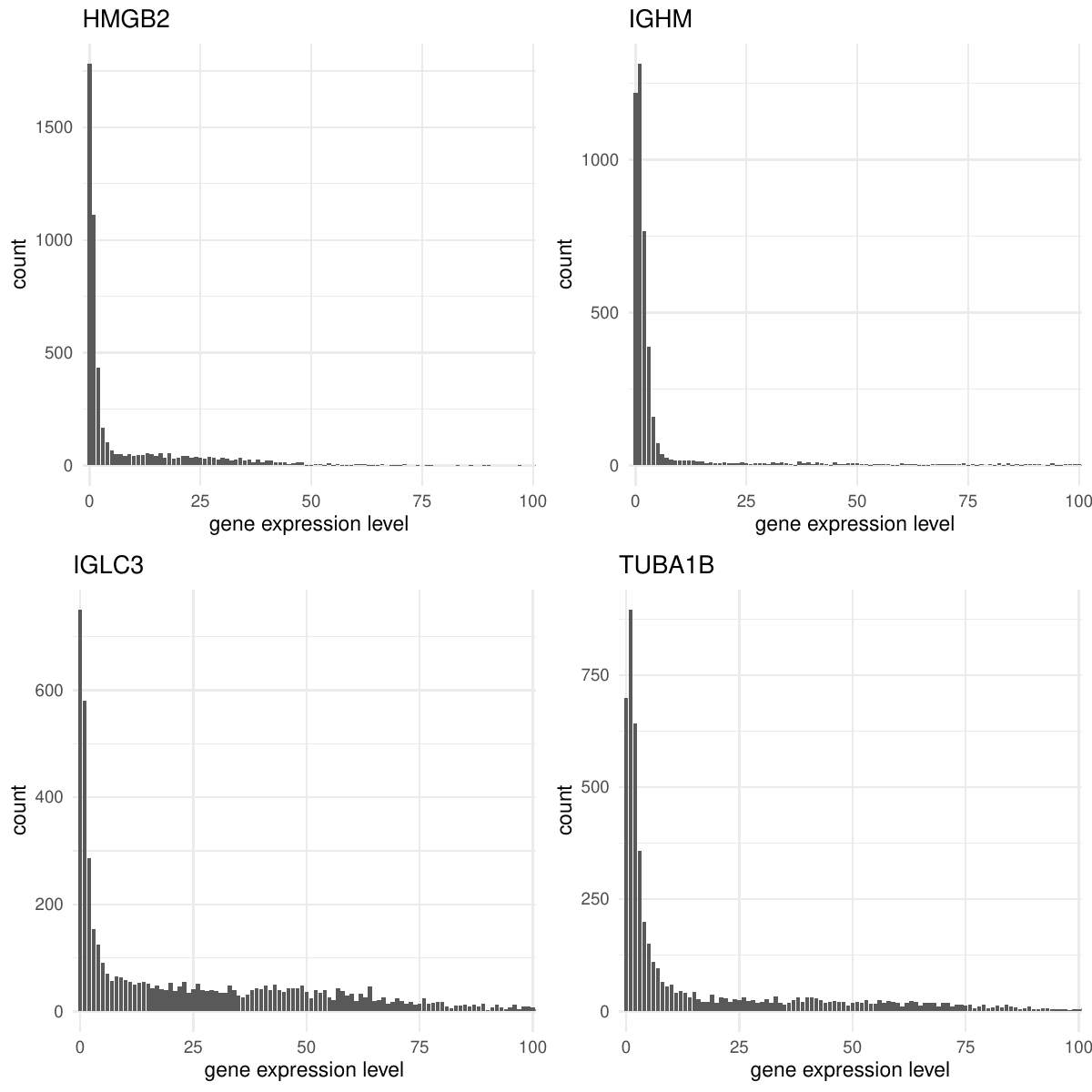}
\caption{\label{fig:genes_selected} Histograms of the gene expression levels of \texttt{HMGB2}, \texttt{IGHM}, \texttt{IGLC2}, and \texttt{TUBA1B}, chosen from our motivating scRNA-seq dataset. These selected genes show a significant inflation of low counts, particularly at counts of 0, 1, and potentially 2.}
\end{figure}

Single-cell sequencing measures the nucleic acid sequence information in individual cells, revealing a high resolution of cellular differences to facilitate our understanding of individual cell functions. Single-cell data analysis is revolutionizing the landscape of whole-organism science, enabling unbiased identification of previously undiscovered molecular heterogeneity at the cellular level \citep{Haque2017-mv,Hwang2018-jh}. Despite this promise, a prominent challenge in analyzing single-cell data is the prevalence of dropouts, where sequencing measurements of cells with limited sequencing information are reported as lower than their true levels due to the low sequencing depth. These dropouts result in an abundance of zero and near-zero measurements in the data, leading to difficulties in the analysis of single-cell data \citep{Lin2017-cp,Ran2020-aj}. This is the case for our motivating dataset that has a high frequency of low counts (0, 1, and possibly 2) as illustrated in Figure \ref{fig:genes_selected} for a few genes. In addition to the prevalence of low counts, the single-cell sequencing data are highly skewed. Both low counts and skewness lead to misleading inference when traditional Gaussian factor models are applied \citep{Pierson2015-pm}.

Several extensions of factor models, which go beyond the traditional Gaussianity assumption, have been proposed. In the realm of parametric models,
\cite{Wedel2003-as} developed Poisson factor models that can provide a parsimonious representation of multivariate dependencies in count data, later extended to dynamic count data by \cite{Acharya2015-un}.
\cite{Sammel2002-iy,Dunson2000-jn} proposed a latent variable model for the factor analysis of mixed data by incorporating shared latent factors in separate generalized linear models for each observed variable, assuming the marginal distributions are in the exponential family. Parametric models, however, might struggle to capture complex multivariate dependencies in real data due to their strict distributional assumptions, which are often proven inadequate in real-world scenarios \citep{Murray2013-kq}. This limitation has prompted the development of semiparametric latent factor models.
\cite{Song2010-bd} assumed flexible nonparametric error distributions instead of Gaussian error distributions, while \cite{Yang2010-ww} proposed a wide class of semiparametric structural equation models that accommodate an unknown distribution for factor scores. 
\cite{Murray2013-kq} proposed a semiparametric Bayesian Gaussian copula factor model for the analysis of mixed data. However, these parametric and semiparametric models were not designed for single-cell data, and thus cannot fully handle the data skewness and heavy low-count inflation.

Factor models for single-cell sequencing data have also been proposed, with a focus on incorporating zero inflation.
\cite{Lee2013-pg} developed a Poisson factor model with offsets that explicitly
takes into account the abundance of zero reads in single-cell data and automatically incorporates sample normalization.
\cite{Cao2017-ze} proposed a Poisson-multinomial model to address the high variation that arises due to excessive zeros. 
To explicitly account for the presence of dropouts in single-cell data, zero-inflated models have been utilized to extend factor models to address sequencing read data with zero-inflation \citep{Pierson2015-pm,B_Sohn2018-aw,Xu2021-mv}. 
While these methods have successfully tackled the zero-inflation characteristic of single-cell sequencing data,  their limitation lies in the inability to account for the inflation of near-zero values, which should also be regarded as dropout events. Moreover, these methods heavily depend on parametric assumptions, potentially limiting their ability to address the high skewness inherent in single-cell data. Therefore, a notable gap remains in the availability of a dimension reduction approach that not only addresses the inflation of low counts in single-cell sequencing data but also demonstrates the flexibility necessary to deal with the high skewness prevalent in such datasets.

In this paper, we propose a segmented Gaussian copula factor model (scFM) to learn low-dimensional latent representations that capture dependencies of single-cell sequencing data with excessive low counts. Unlike existing factor models for single-cell data, the proposed approach leverages the flexibility of copulas, enabling the accommodation of heavy-tailed distributions while effectively addressing the issue of inflated low counts via segmentation. 
We also introduce a Bayesian framework that utilizes a column-wise Dirichlet-Laplace prior \citep{Bhattacharya2015-pm} for factor loadings to accommodate the uncertainty in the number of latent factors and allow for automatic selection of the number of significant latent factors. Notably, our Bayesian approach effectively resolves the identifiability issue inherent in factor models due to the invariance to rotations of the factor loadings matrix \citep{Anderson1955-ep}. Unlike conventional approaches that impose the lower triangular constraint on the factor loadings matrix \citep{Geweke1995-oc,Aguilar2000-hr}, which makes the factor scores dependent on the ordering of variables, our method circumvents this need for subjective decisions on variable ordering. By resolving the identifiability issue without resorting to such triangular constraints, our approach ensures that inference results and their interpretation remain independent of variable ordering and facilitates consistent results in downstream analyses.
Lastly, an efficient data-augmented Gibbs sampler is developed for posterior inference on the proposed model, and the utility of the proposed method is empirically demonstrated through both simulation studies and a real data application. The major contributions of this paper are four-fold:
\begin{itemize}
    \item We introduce scFM, a novel approach for factor analysis on single-cell sequencing data characterized by excessively low counts. The proposed copula framework is adept at addressing the inflation of low counts while accommodating heavy-tailed distributions.
    \item We propose a Bayesian framework that incorporates the uncertainty in the true number of latent factors and automatically estimates the number of factors.
    \item The proposed approach effectively resolves the identifiability issue in factor models, deviating from conventional methods that impose triangular constraints on factor loadings, thereby eliminating the dependency on variable ordering and enhancing the interpretability of latent factors.
    \item We develop an efficient data-augmented Gibbs sampler for posterior inference.
\end{itemize}


\section{Bayesian segmented Gaussian copula factor model}
\label{sec:model}

In this section, we introduce a Bayesian semiparametric approach, scFM, for modeling dependencies in skewed count data with an excessive number of low counts.

\subsection{Segmented Gaussian copula factor model}
\label{sec:tgcfm}

Let $\xb = (x_1, \ldots, x_p)^\top$ denote $p$ random variables of interest, i.e., read counts of an individual cell in single-cell sequencing data. In order to account for an excessive number of low counts, including zeros, 
we propose the segmented Gaussian copula model. We begin by reviewing the standard Gaussian copula model developed by \cite{Liu2009-vm}.

\begin{definition}[Gaussian copula model]\label{def:GCM}
A random vector $\xb = (x_1, \ldots, x_p)^\top$ satisfies the Gaussian copula model (also known as the nonparanormal model), if there exist strictly increasing transformations $\{f_j\}_{j=1}^p$ such that $(z_1, \ldots, z_p)^\top := (f_1(x_1), \ldots, f_p(x_p))^\top \sim \N(\bm{0}, \Omegab)$, where $\Omegab$ is a correlation matrix. We denote this by $\xb \sim \NPN(\bm{0}, \Omegab, \fb)$.
\end{definition}

\begin{figure}
\centering
\includegraphics[width=0.9\linewidth]{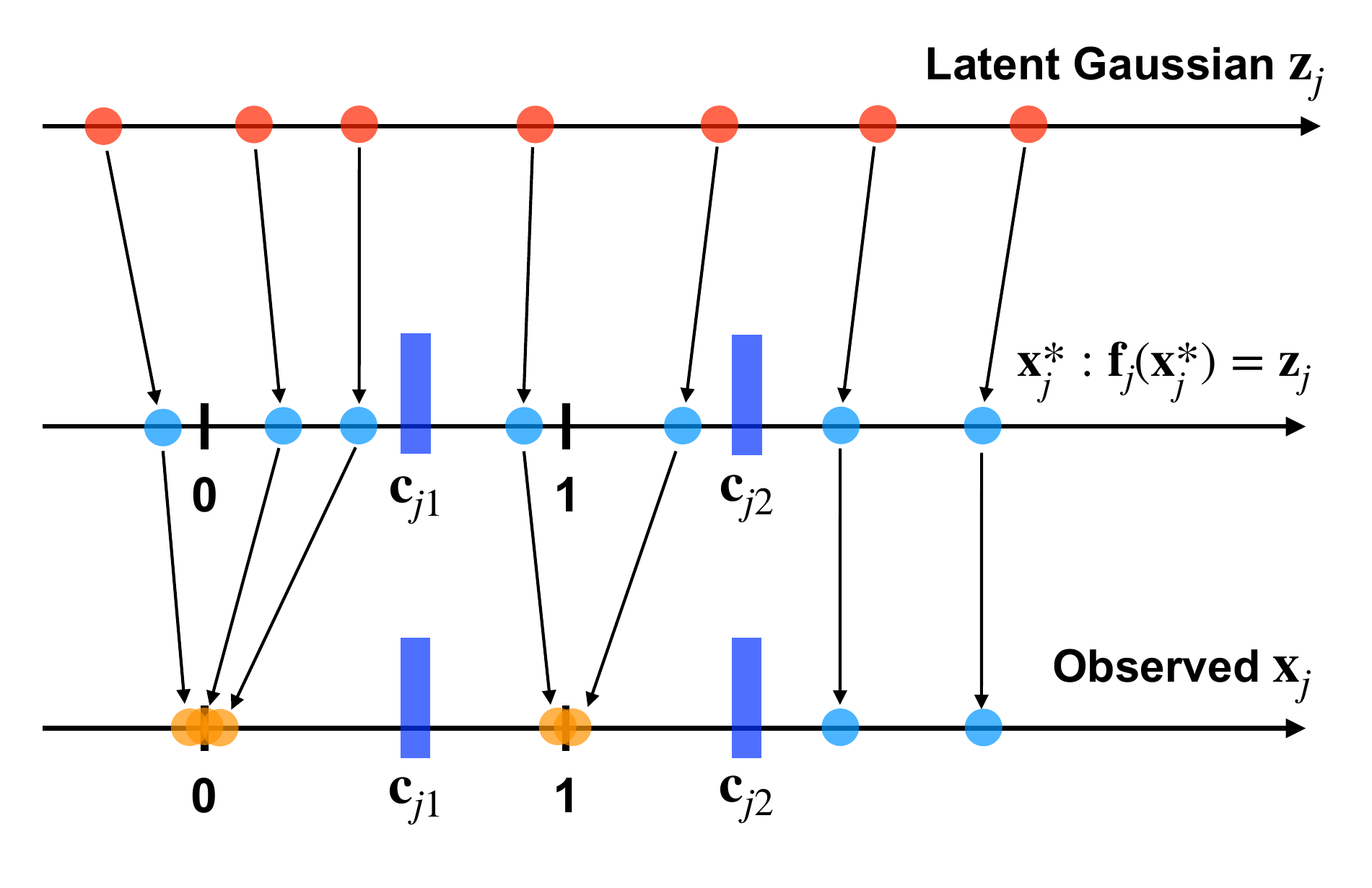}
\caption{\label{fig:SGC} A schematic of the segmented Gaussian copula model when $m = 1$. In the segmented Gaussian copula model, observed variables $\xb = (x_1, \ldots, x_p)^\top$ are generated through the following procedure: initially, a latent random vector $\xb^* = (x_1^*, \ldots, x_p^*)^\top$ is generated from the latent Gaussian vector $\zb = (z_1, \ldots, z_p)^\top$ through the monotone transformations $f_j$. The latent random vector $\xb^*$ serves as an underlying source for the observed data, and a segmentation process is applied to $\xb^*$ to generate the observed $\xb$. When $m=1$, each latent random variable $x_j^*$ is segmented into three intervals $(-\infty, c_{j1}]$, $(c_{j1}, c_{j2}]$, and $(c_{j2}, \infty)$. If $x_j^*$ falls into either one of the first two intervals $(-\infty, c_{j1}]$ or $(c_{j1}, c_{j2}]$, our observation will be either $x_j = 0$ or $1$. On the other hand, if 
$x_j^*$ is greater than $c_{j2}$, we will directly observe the latent source $x_j^*$ (i.e., $x_j = x_j^*)$.}
\end{figure}

Our proposed segmented Gaussian copula model accounts for dropouts (low counts) by associating them with segments of underlying latent continuous values. Let $m \in \{0, 1, \ldots\}$ be a hyperparameter determining the maximum low count being inflated in the observed data.
Figure \ref{fig:SGC} illustrates a schematic of the proposed segmented Gaussian copula model when $m = 1$, and the formal definition for any $m$ follows below.

\begin{definition}[Segmented Gaussian copula model]\label{def:SGCM}
A random vector $\xb = (x_1, \ldots, x_p)^\top$ satisfies the segmented Gaussian copula model, if there exists a latent random vector 
$\xb^* \sim \NPN(\bm{0}, \Omegab, \fb)$ 
and strictly increasing constants $-\infty=c_{j0} < c_{j1} < \cdots < c_{jd} < c_{j,m+1} < \infty$ such that
\begin{align}\label{eq:SGCM}
x_j = \sum_{d=0}^m 1(c_{jd} < x_j^* \le c_{j,d+1}) d + 1(x_j^* > c_{j,m+1}) x_j^*. 
\end{align}
\end{definition}

As illustrated in Figure~\ref{fig:SGC} and formally introduced in Definiton \ref{def:SGCM}, the observed variables $\xb$ are generated through a two-stage process.
Initially, a latent random vector $\xb^*$ is generated from the Gaussian copula model, serving as an underlying source for the observed data. Subsequently, the observed data $\xb$ are derived from the latent random vector $\xb^*$ via a segmentation process. Specifically, the latent random variable $x_j^*$ is segmented into multiple intervals $(c_{jd}, c_{j,d+1}]$, $d=0, \ldots, m$, and $(c_{j,m+1}, \infty)$. If $x_j^*$ exceeds a specific threshold $c_{j, m+1}$, corresponding to the interval $(c_{j,m+1}, \infty)$, the latent source $x_j^*$ is directly observed (i.e., $x_j = x_j^*)$. Otherwise, if $x_j^* \in (c_{jd}, c_{j,d+1}]$ where $d \in \{0, 1, \ldots, m\}$, it results in the observation of a low count value $x_j = d$. Hence, $\cb = (c_{j0}, \ldots, c_{j,m+1})$ represents unknown thresholds that truncate the latent random variable $x_j^*$ to near-zero counts for the observed variable $x_j$. Additionally, given that the transformations $f_j$ are strictly increasing, \eqref{eq:SGCM} can be equivalently expressed using the latent Gaussian vector $\zb = (f_1(x_1^*), \ldots f_p(x_p^*))^\top$,
\begin{align}\label{eq:deltas}
x_j = \sum_{d=0}^m 1(\delta_{jd} < z_j \le \delta_{j,d+1}) d + 1(z_j > \delta_{j,m+1}) x_j^*, 
\end{align}
where $\delta_{jd} = f_j(c_{jd})$ is the latent Gaussian level threshold corresponding to $c_{jd}$. 

When $m=0$, the proposed segmented Gaussian copula model is reduced to the truncated Gaussian copula model of \cite{Yoon2020-qv}, which focuses on addressing zero inflation. When $m > 0$, the segmented Gaussian copula model is capable of accommodating the inflation of low counts besides zero. For example, if $m = 1$, the latent variable $x_j^*$ is segmented into three intervals $(c_{j0}, c_{j1}], (c_{j1}, c_{j2}]$, and $(c_{j2}, \infty)$. If $x_j^*$ falls into either one of the first two intervals $(c_{j0}, c_{j1}]$ and $(c_{j1}, c_{j2}]$, our observation will be either $x_j = 0$ or $1$. This allows for a flexible representation of zero and one counts in the observed data. 

Next, we extend the proposed segmented Gaussian copula to scFM to capture underlying dependencies among observed genes in single-cell sequencing data while simultaneously addressing the issue of inflated near-zero counts. We adopt a low-rank decomposition of  the copula correlation matrix $\Omegab$ so that the dependence among the latent variables can be explained by low-dimensional latent factors.

\begin{definition}[Segmented Gaussian copula factor model]\label{def:SGCFM}
We say that a random vector $\xb = (x_1, \ldots, x_p)^\top$ follows the segmented Gaussian factor model (scFM), if $\xb$ satisfies the segmented Gaussian copula model in Definition \ref{def:SGCM} and the latent Gaussian vector $\zb = (f_1(x_1^*), \ldots, f_p(x_p^*))^\top$ is defined by a probabilistic factor model,
\begin{align}\label{eq:factor-model}
\Psib^{1/2} \zb = \Lambdab \ub + \varepsilonb,
\end{align}
where $\Psib^{1/2}$ is a diagonal matrix that adjusts the scale of $\zb$ to ensure that its covariance matrix $\Omegab$ is a correlation matrix,
$\Lambdab$ is a $p \times k$ matrix of factor loadings, $\ub \sim \N(\bm{0}, \Ib)$ is a $k \times 1$ latent factor scores, and $\varepsilonb \sim \N(\bm{0}, \Sigmab)$ is an idiosyncratic Gaussian noise independent of $\ub$ with $\Sigmab = \diag
\{\sigma_j^2\}_{j=1}^{p}$. 
\end{definition}

Note that the covariance matrix $\bm{\Omega}$ of the latent Gaussian $\zb$ is constrained to be a correlation matrix for identifiability \citep{Liu2009-vm}, and thus
$\Omegab = \Psib^{-1/2} (\Lambdab \Lambdab^\top + \Sigmab) \Psib^{-1/2}$ implies 
$\Psib = \diag\{\lambdab_{j\cdot} ^\top \lambdab_{j\cdot} + \sigma_j^2 \}_{j=1}^{p}$, 
where $\lambdab_{j\cdot}$ is the $j$-th row of the factor loading matrix $\Lambdab$. 
In practice, it is typically assumed that the number of latent factors $k$ is much smaller than the number of observed variables $p$, leading to a low-rank characterization of $\Omegab$. Therefore, once the factor scores $\Ub$ are estimated, the dimensionality of the dataset can be effectively reduced from $p$ to a lower dimension $k$ $(\ll p)$, which facilitates downstream analyses. 
A fundamental challenge here is that the number of latent factors $k$ is typically unknown \emph{a priori}. Hence, the selection of an appropriate $k$ is of great significance. 
In the next subsection, we introduce a meticulously selected prior distribution for the factor loadings that enables the automatic selection of the number of latent factors as well as facilitates the identifiability of latent factors.

\subsection{Prior specification}
\label{sec:prior}

We propose to use a column-wise Dirichlet-Laplace (DL, \citealt{Bhattacharya2015-pm}) prior for the factor loading matrix $\Lambdab = (\lambda_{jh})$. 
The column-wise DL prior enables simultaneous shrinkage of all the entries of a column, thereby effectively zeroing out unnecessary latent factors.
Specifically, we assume, for $j = 1, \ldots, p$ and $h = 1, \ldots, k_{\text{max}}$,
\begin{align}\label{eq:DL}
\lambda_{jh} & \sim \mbox{DE}(\tau \phi_{h}), \nonumber \\
\tau & \sim \mbox{Gamma}(pk_{\text{max}} \alpha, 1/2), \nonumber \\
\phib & \sim \mbox{Dir}(\alpha, \ldots, \alpha),
\end{align}
where $\phib = (\phi_{1}, \ldots, \phi_{k_{\text{max}}})^\top$, 
$\mbox{DE}(\tau\phi_{h})$ denotes the Laplace or double-exponential distribution with scale parameter $\tau\phi_{h}$, $\mbox{Dir}(\alpha, \ldots, \alpha)$ denotes the Dirichlet distribution with concentration parameter $\alpha$ which is considered as a hyperparameter.

The global scale $\tau$ controls the overall shrinkage towards zero, while the column-specific local scale $\phi_{h}$ allows each column of the factor loading matrix $\Lambdab$ to deviate away from zero.
Here $k_{\text{max}}$ represents the maximum allowable number of latent factors. Since DL prior allows for aggressive shrinkage of all entries of a column towards zero, the effective number of factors can be much smaller than $k_{\text{max}}$, and is determined automatically; see Section~\ref{sec:estim_k*}.

Intuitively, the class of column-wise DL priors approximates a spike-and-slab prior \citep{George1997-ey} assigned to the factor loadings at the column level by a continuous density concentrated near zero with a heavy tail.
A distinguishing characteristic of the DL prior, in contrast to commonly used shrinkage priors like Laplace priors \citep{Park2008-cu}, is the presence of a singularity at zero in its distribution while still maintaining exponential tails. Under the proposed column-wise DL priors, the singularity at zero facilitates the concentration of most of the columns of $\Lambdab$ around zero, while the exponential tails ensure little shrinkage of a small number of significant factors.
See \cite{Bhattacharya2015-pm} for the theoretical properties of the DL prior. 

In addition to the adaptive shrinkage, imposing DL priors for the factor loadings also resolves the identifiability issue of factor models.
In the Bayesian framework, a typical approach for factor analysis is assigning Gaussian priors to the factor loadings \citep{Lopes2004-vt}. Although this approach allows for a conditionally conjugate model, facilitating posterior inference through standard Gibbs sampling,  it fails to address the invariance of factor models to orthogonal transformations as $\Lambdab \ub=(\Lambdab \Pb) (\Pb^\top\ub):=\Lambdab^* \ub^*$ for any orthogonal matrix $\Pb$ \citep{Anderson1955-ep}.
A popular technique to deal with this invariance property is imposing conditions that remove invariance to orthogonal transformation, e.g., the lower-triangular constraint on $\Lambdab$ with an assumption of strictly positive diagonal elements \citep{Geweke1995-oc,Aguilar2000-hr}. 
While this approach is effective for estimating the covariance matrix, the constraint introduces order-dependence in estimating factor loadings, making the interpretation of latent factors sensitive to the ordering of observed variables. Consequently, changes in variable ordering can influence downstream analyses, such as clustering, thereby compromising the reproducibility of their results. 
In contrast, our approach, which assigns the column-wise DL priors to the factor loadings, breaks this invariance in the posterior distribution without necessitating the lower-triangular constraint on the factor loadings matrix. Our factor model \eqref{eq:factor-model} with non-Gaussian DL prior can be viewed as a noisy independent component analysis (ICA) model, where the columns of $\Lambdab$ are the hidden sources, $\ub$ is the mixing matrix, and $\epsilonb$ is the noise. Since the sources $\Lambdab$ are non-Gaussian, 
the theory of noisy ICA model ensures identifiability of our model parameters $\Lambdab$ and $\ub$ (up to column permutations and sign changes)
\citep{Comon1994-bu, Hyvarinen2004-wh}. 
In summary, our approach alleviates the burden of subjective decisions regarding the ordering of observed variables, which often introduces ambiguity in conventional factor models. As a result, our approach achieves greater consistency in the interpretation of latent factors, and contributes to the robustness and reliability of downstream analyses.

For the Gaussian level thresholds $\deltab=(\delta_{j0},\ldots,\delta_{j,d})^{\top}$ in \eqref{eq:deltas}, we assume an improper uniform prior with the following monotonicity constraint, 
\begin{align*}
p(\delta_{j1}, \ldots, \delta_{j,m+1}) & \propto 1(\delta_{j1} \le \cdots \le \delta_{j,m+1}) \quad \mbox{for} \quad j = 1, \ldots, p.
\end{align*}
For the error variances $\sigma_j^2$, we assign a conditionally conjugate inverse-gamma prior $\sigma_j^2 \sim \IG(a_\sigma, b_\sigma)$.

\section{Posterior inference}
\label{sec:infer}

In this section, we detail the posterior inference procedure for the proposed scFM model, focusing on the case where $m=1$, which involves inflation of zero and one values. Extending the procedure to accommodate other values of $m$ is straightforward.
Let $\xb_i \in \R^p$, $i = 1, \ldots, n$, be samples from the proposed scFM, and $\zb_i \in \R^p$ and $\ub_i \in \R^k$, $i = 1, \ldots, n$, be the corresponding latent Gaussian vectors and latent factor scores, respectively. We denote the sub-vectors of $\xb_i$ containing zero and one values by $\xb_i^0$ and $\xb_i^1$, respectively. The observed sub-vectors of $\xb_i$ (containing values greater than one) are denoted as $\xb_i^o$. Likewise, we denote by $\zb_i^0, \zb_i^1$, and $\zb_i^o$ the corresponding latent Gaussian sub-vectors.


\subsection{Copula transformations}

The latent Gaussian variables corresponding to the observed data $\xb_i^o$ are defined by $z_{ij}^o = f_j(x_{ij}^o)$.  We will first compute ``observed pseudodata'' $\hat{z}_{ij}^o$ by estimating $f_j$ via the empirical cumulative distribution function (cdf).


Let $F_j$ be the (monotonically increasing) marginal cdf of the $j$-th latent random variable $x_j^*$ and $\Phi$ be the cdf of standard Gaussian. 
Then, we observe
$$
F_j(a) = P(x_j^* \le a) = P(z_j \le f_j(a)) = \Phi(f_j(a)),
$$
which implies that $f_j = \Phi^{-1} \circ F_j$.
Since $F_j$ is unknown, we replace it with its empirical estimate \citep{Chris_A_J_Klaassen1997-sk},
\begin{align}\label{eq:empir-cdf}
\hat{F}_{j}(x)=\frac{n}{n+1}\sum_{i=1}^{n} \frac{1}{n} 1(x_{ij} \leq x).
\end{align}
Here, the constant term $n / (n+1)$ is necessary to restrict $\hat{F}_j(x) < 1$ to make $\hat{z}_{ij}^o$ finite. 
Given $\hat{F}_{j}$, we compute the observed pseudodata $\hat{z}_{ij}^o = \hat{f}_j(x_{ij}^o) = \Phi^{-1} \circ \hat{F}_j(x_{ij}^o)$.

\subsection{Data-augmented Gibbs sampling}

Given the observed pseudodata, we employ an MCMC algorithm to draw posterior samples of the model parameters.
Let $ \hat{\Zb}^o = \{\hat{\zb}_i^o\}_{i=1}^n, \Zb^{\leq 1} = \{(\zb_i^0, \zb_i^1)\}_{i=1}^n$, and $\Ub = \{\ub_i\}_{i=1}^n$.
Given $\hat{\Zb}^o$, we use MCMC to draw posterior samples of $\Zb^{\leq 1}, \deltab, \Sigmab, \Ub, \Lambdab, \phib$, and $\tau$. 
In particular, we utilize an efficient data-augmented Gibbs sampler, which is derived by a data-augmented representation of the proposed column-wise DL priors. 
Introducing auxiliary variables $\Xib = (\xi_{jh})_{j,h}$, the DL prior \eqref{eq:DL} can be equivalently represented as
\begin{align*}
\lambda_{jh} & \sim \mbox{N}(0, \xi_{jh} \tau^2 \phi_{h}^2), \\
\xi_{jh} & \sim \mbox{Exp}(1 / 2), \\
\phib & \sim \mbox{Dir}(\alpha, \ldots, \alpha), \\
\tau & \sim \mbox{Gamma}(pk_{\text{max}} \alpha, 1/2),
\end{align*}
which facilitates Gibbs sampling by enabling the derivation of full conditionals  for all  parameters.
Our implementation of the Gibbs sampler leverages marginalization and blocking to reduce auto-correlation in the posterior samples. 
It is given that, upon conditioning on the latent Gaussian vectors $\Zb$, the thresholds $\deltab$ become independent of the other parameters, and similarly, conditioning on the factor loadings $\Lambdab$, the hyperparameters $(\Xib, \tau, \phib)$ are independent from the remaining parameters.
Utilizing these conditional independences, our Gibbs sampler cycles through the following draws:
(i) $\Zb^{\leq 1} | \deltab, \Sigmab, \Ub, \Lambdab, \hat{\Zb}^o$, (ii) $\deltab | \Zb^{\leq 1}, \hat{\Zb}^o$, (iii) $\Sigmab | \Ub, \Lambdab,  \Zb^{\leq 1}, \hat{\Zb}^o$, (iv) $\Ub | \Sigmab, \Lambdab, \Zb^{\leq 1}, \hat{\Zb}^o$, (v) $\Lambdab | \Sigmab, \Ub,  \Xib, \tau, \phib, \Zb^{\leq 1}, \hat{\Zb}^o$, and (vi) $\Xib, \tau, \phib | \Lambdab$.

\underline{Sample $\Zb^{\leq 1} | \deltab, \Sigmab, \Ub, \Lambdab, \hat{\Zb}^o$.} We draw $(\zb_i^0, \zb_i^1)$ from a truncated multivariate Gaussian distribution,
\begin{align}\label{eq:gibbs-z}
p(\zb_i^0, \zb_i^1 | \deltab, \Sigmab, \Ub, \Lambdab, \hat{\Zb}^o ) \propto \N(\zb_i^0, \zb_i^1; \mub_i, \Cb_i) 1(\deltab_{i, 0}^0 < \zb_i^0 \le \deltab_{i, 1}^0) 1(\deltab_{i, 1}^1 < \zb_i^1 \le \deltab_{i, 2}^1),
\end{align}
where $\mub_{i}$ and $\Cb_{i}$ are the conditional mean and conditional covariance matrix of $(\zb_i^0, \zb_i^1)$ given $\zb_i^o = \hat{\zb}_i^o$ and $(\Sigmab, \Ub, \Lambdab)$. Since $\cov(z_{ij}, z_{ij'} | \Sigmab, \Ub, \Lambdab) = 0$ for $j \ne j'$,  $\mub_i$ and $\Cb_i$ are the sub-vector and sub-matrix of $\E(\zb_i | \Sigmab, \Ub, \Lambdab) = \Psib^{-1/2} \Lambdab \ub_i$ and $\Var(\zb_i | \Sigmab, \Ub, \Lambdab) = \Psib^{-1/2} \Sigmab \Psib^{-1/2}$ corresponding to the variables $(\zb_i^0, \zb_i^1)$, respectively. In \eqref{eq:gibbs-z}, 
the vector $\deltab_{i, d}^a = \{\delta_{jd}: x_{ij} = a, j = 1, \ldots, p\}$ consists of the thresholds $\delta_{jd}$ for the $i$-th observation, specifically corresponding to those variables $x_{ij}$ that equal $a$. The subscript $d$ signifies whether the threshold represents a lower bound $(d = a)$ or an upper bound $(d = a+1)$, respectively.

\underline{Sample $\deltab | \Zb^{\leq 1}, \hat{\Zb}^o$.} The conditional posterior of $\deltab | \Zb^{\leq 1}, \hat{\Zb}^o$ is proportional to the $n$-product of indicator functions in \eqref{eq:gibbs-z}, and thus Step (ii) can be done by sampling $(\delta_{j1}, \delta_{j2}), j = 1, \ldots, p$, independently  from the uniform full conditional distribution
\begin{align*}
\textstyle
p(\delta_{j1}, \delta_{j2} | \Zb^{\leq 1}, \hat{\Zb}^o ) \propto 1(  \max_i z_{ij}^0 < \delta_{j1} < \min_i z_{ij}^1) ~1(\max_i z_{ij}^1 < \delta_{j2} < \min_i z_{ij}^o).
\end{align*}

\underline{Sample $\Sigmab | \Ub, \Lambdab,  \Zb^{\leq 1}, \hat{\Zb}^o$.} We draw $\sigma_j^2$ from $\sigma_j^2 \sim \IG( a_{\sigma^2} + n/2, b_{\sigma^2} +  \sum_{i=1}^n(z_{ij} - \lambdab_j^\top \ub_i))$. 

\underline{Sample $\Ub | \Sigmab, \Lambdab, \Zb^{\leq 1}, \hat{\Zb}^o$.} We draw $\ub_i \sim \N(\mb_i, \Vb_i)$ where $\mb_i = \Vb_i (\Lambdab^\top \Sigmab^{-1} \zb_i)$ and $\Vb_i = (\Lambdab^\top \Sigmab^{-1} \Lambdab + \Ib_{k_{\text{max}}})^{-1}$.

\underline{Sample $\Lambdab | \Sigmab, \Ub,  \Xib, \tau, \phib, \Zb^{\leq 1}, \hat{\Zb}^o$.}
We draw each row of $\Lambdab$ from $\lambdab_j \sim \N(\etab_j, \Wb_j)$, where $\etab_j = \Wb_j (\sigma_j^{-2} \sum_{i=1}^n z_{ij} \ub_i)$ and $\Wb_j = (\sigma_j^{-2} \sum_{i=1}^n \ub_i \ub_i^\top + \Db^{-1})^{-1}$ with 
$\Db = \diag\{ \xi_{jh} \tau^2 \phi_{jh}^1\}_{h=1}^{k_{\text{max}}}$.

\underline{Sample $\Xib, \tau, \phib | \Lambdab$.} 
Since 
\begin{align*}
p(\Xib, \tau, \phib  | \Lambda) = p(\phib | \Lambdab) p(\tau | \phib, \Lambdab) p(\Xib | \tau, \phib, \Lambdab),
\end{align*}
we sequentially sample from $\phib | \Lambdab$,  $\tau | \phib, \Lambdab$, and $\Xib | \tau, \phib, \Lambdab$.
For sampling $\phib | \Lambdab$, we follow the strategy in \cite{Bhattacharya2015-pm}. Specifically, let $\giG(\kappa, \rho, \chi)$ denote a three-parameter generalized inverse Gaussian distribution with the density $p(y) \propto y^{\kappa-1} e^{-0.5 (\rho y + \chi / y)}$ for $y > 0$. We draw $R_{11}, \ldots, R_{pk_{\text{max}}}$ independently, $R_{jh} \sim \mbox{giG}(\alpha-1, 1, 2 | \lambda_{jh}| )$, and set $\phi_{jh} = R_{jh}/ R$ with $R = \sum_{j, h} R_{jh}$. Given $\phib$, we then sample $\tau | \phib, \Lambdab$ from  $\mbox{giG}(p k_{\text{max}} (\alpha - 1), 1, 2 \sum_{j, h} | \lambda_{jh} |/ \phi_{jh} )$. 
Lastly, we sample $\Xib | \phib, \tau, \thetab$ by drawing $\xi_{jh}$ independently from $\mbox{giG}(1/2, 1, \lambda_{jh}^2 / (\tau^2 \phi_{jh}^2))$.
An outline of our data-augmented Gibbs sampler is provided in Algorithm \ref{alg:mcmc}.

\begin{algorithm}[!t]
\caption{Outline of the proposed data-augmented Gibbs sampler.}\label{alg:mcmc}
\begin{algorithmic}[1]
\State \textbf{Input:} $\hat{\zb}_1^o, \ldots, \hat{\zb}_n^o$, where $\hat{z}_{ij}^o = \hat{f}_j(x_{ij}^o)$, and the maximum number $k_{\text{max}}$ of latent factors.
\State Initialize $\Zb^{\leq 1}, \deltab, \Sigmab, \Ub, \Lambdab, \phib$, and $\tau$.
\For{$t = 1, \ldots, T$}
\State Sample truncated Gaussian vectors $(\zb_i^0, \zb_i^1) \sim p(\zb_i^0, \zb_i^1 | \deltab, \Sigmab, \Ub, \Lambdab, \hat{\Zb}^o ), i = 1, \ldots, n$.
\State Sample threshold parameters $(\delta_{j0}, \delta_{j1}) \sim p(\delta_{j0}, \delta_{j1} | \Zb^{\leq 1}, \hat{\Zb}^o ), j = 1, \ldots, p$.
\State Sample noise variances $\sigma_j^2 \sim \IG( a_{\sigma^2} + n/2, b_{\sigma^2} +  \sum_{i=1}^n(z_{ij} - \lambdab_j^\top \ub_i)), j =1, \ldots, p$.
\State Sample factor scores $\ub_i \sim \N(\mb_i, \Vb_i), i = 1, \ldots, n$.
\State Sample factor loadings $\lambdab_j \sim \N(\etab_j, \Wb_j), j = 1, \ldots, p$.
\State Sample $R_{jh} \sim \mbox{giG}(\alpha-1, 1, 2 | \lambda_{jh}| ), j = 1, \ldots, p; h = 1, \ldots, k_{\text{max}}$, and set local scale parameters for the column-wise DL prior $\phi_{jh} = R_{jh}/ R$ with $R = \sum_{j, h} R_{jh}$.
\State Sample global scale parameter for the column-wise DL prior $\tau \sim \mbox{giG}(p k_{\text{max}} (\alpha - 1), 1, 2 \sum_{j, h} | \lambda_{jh} |/ \phi_{jh} )$.
\State Sample auxiliary variables $\xi_{jh} \sim \mbox{giG}(1/2, 1, \lambda_{jh}^2 / (\tau^2 \phi_{jh}^2))$, $j = 1, \ldots, p$; $h = 1, \ldots, k_{\text{max}}$.
\EndFor
\end{algorithmic}
\end{algorithm}

\subsection{Estimation of the number of significant latent factors}\label{sec:estim_k*}

In the proposed scFM, $k_{\text{max}}$ is the maximum allowable number of latent factors, and the number of significant factors, denoted by $\hat{k}\leq k_{\text{max}}$, is determined by the proposed column-wise DL prior.
The column-wise DL prior induces a shrinkage effect on factor loadings associated with irrelevant latent factors that lack support from the data, thereby aiding in the identification of $\hat{k}$. More specifically, our DL prior promotes the concentration of the majority of columns in the factor loadings matrix $\Lambdab$ around zero, except for a few columns corresponding to a limited number of significant factors. This implies the existence of potentially two clusters of columns in $\Lambdab$: one cluster closely centered around zero, indicating insignificant factor, and another cluster distant from zero, indicating significant factors.

Consequently, we suggest a simple automated approach for selecting $\hat{k}$. At each MCMC iteration, we cluster the columns of the factor loadings matrix $\Lambdab$ into two clusters using k-means, and estimate the number of significant factors by the size of the cluster that exhibits greater deviance from zero. The  number of significant factors $\hat{k}$ is then estimated by taking the mode over all the MCMC iterations. Upon obtaing the estimated $\hat{k}$, we identify the significant factors by selecting the latent factors associated with the $\hat{k}$ largest columns (based on $\ell_2$-norm) of the posterior mean of $\Lambdab$.

\section{Simulation study}
\label{sec:sim}

We empirically evaluate the performance of the proposed scFM with synthetic data. We compare the proposed method against other factor models for single-cell sequencing data: the zero-inflated factor analysis (ZIFA, \citealp{Pierson2015-pm}) and the zero-inflated Poisson factor analysis (ZIPFA, \citealp{Xu2021-mv}). We also compare scFM with the Bayesian Gaussian copula factor model (CopulaFM, \citealp{Murray2013-kq}), which is a  flexible semiparametric factor model for mixed data.

To generate synthetic data with marginal distributions matching the empirical marginal distributions of our motivating scRNA-seq data, we follow the ideas in  \cite{Chung2022-at}. 
Specifically, we generate the latent Gaussian vectors $\zb_i$ from \eqref{eq:factor-model} with the true factor loadings sampled from the Laplace distribution $\lambda_{jk} \sim \mbox{DE}(1)$ and the true error variances sampled from the uniform distribution $\sigma_j^2 \sim U(0.3, 1)$. To obtain the observed data $\xb_i$ from the latent Gaussian $\zb_i$, we use the empirical cdfs of the motivating scRNA-seq dataset in Section \ref{sec:reald}. First, we select $p$ genes that exhibit the highest cell-to-cell variations and use the empirical cdf of each selected gene as true $F_j$. The observed data  $\xb_i$ are then obtained by $x_{ij} = F_j^{-} \circ \Phi(z_{ij})$, where $F_j^{-}$ is the pseudo inverse of $F_j$. This data generating process allows us to preserve the marginal distributions of the motivating scRNA-seq dataset in our simulated data while introducing the joint dependence structure through the pre-specified $\Lambdab$ and $\Sigmab$. We consider different sample sizes $n \in \{500, 1000, 2000\}$ and different numbers of genes $p \in \{25, 50, 100\}$ while fixing the number of true latent factors $k = 4$. Each scenario is repeated 50 times. The simulated datasets show a significant abundance of zeros ($54\%-59\%$) and ones ($14\%-17\%$).

For the proposed scFM, we set the hyperparameters at $a_\sigma = b_\sigma = 0.1$ and $\alpha = 0.5$. The maximum count value that we consider to be inflated is set to $m = 1$. 
To assess the efficacy of the DL prior in the selection of significant latent factors and its robustness to varying specifications of $k_{\text{max}}$, we examine two distinct values of $k_{\text{max}} \in \{4, 8\}$. The first value ($k_{\text{max}}=4$) corresponds to the true number of factors, while the second value is intentionally overspecified, allowing us to test how well the proposed approach accommodates the uncertainty in the selection of significant latent factors.
For each $k_{\text{max}}$, we run the data-augmented Gibbs sampler proposed in Section \ref{sec:infer} for 10,000 iterations, of which the first 5,000 iterations are discarded as burn-in. Based on the MCMC samples, we calculate the posterior mean of the factor scores and the factor loadings (for $k_{\text{max}}$ factors) to obtain their estimates. For the competitors, ZIFA, ZIPFA, and CopulaFM, we set the number of latent factors to the simulation ground truth (i.e., $k = 4$). Additionally, we apply the $\log(x+1)$ transformation to the simulated data before applying ZIFA in concordance with the data processing pipeline in the corresponding paper.

To evaluate the performance, we follow the approach in \cite{Pierson2015-pm}, which accounts for the rotational invariance of factor model specification in terms of scores and loadings. While the proposed scFM addresses this rotational invariance, we use the approach in \cite{Pierson2015-pm} to be objective in comparison with other methods. The main idea is that while rotation changes the values of scores and loading, it preserves relative distances. Thus instead of comparing the true factor scores $\ub_i$ with estimated factor scores $\hat{\ub}_i$ directly, we will evaluate concordance between corresponding score-induced pairwise distances. Let $\Db\in \R^{n\times n}$ represent a pairwise distance matrix with each element $D_{ii'} = ||\ub_i - \ub_{i'}||_2$ being the Euclidean distance between true factor scores $\ub_i$ and $\ub_{i'}$, and let $\hat \Db$ be the corresponding pairwise distance matrix based on estimated factor scores $\hat{\ub}_i$. We will compute the Spearman correlation between $\Db$ and its estimated counterpart $\hat{\Db}$, with larger correlation values indicating more accurate estimation performance. We will use a similar approach to evaluate the accuracy of estimated factor loadings.


\begin{table}
\caption{Average Spearman correlations between the estimated and true latent distances among factor scores $\ub_i$ and among factor loadings $\lambdab_{j\cdot}$ across different sample sizes and different numbers of genes. The standard error is given within parentheses. The averages and standard errors are computed over 50 replicated datasets.  The proposed scFM has two results, one corresponding to the true $k$ ($k_{\text{max}}=4$) and the other to the overspecified value ($k_{\text{max}} = 8$). The result for CopulaFM when $p = 100$ is omitted due to numerical issues with the R package implementation of CopulaFM in this scenario.}
\label{tab:scores}
\begin{center}
\resizebox{\columnwidth}{!}{%
\begin{tabular}{lcccccc}
\hline 
              & \multicolumn{3}{c}{$p = 50$}                                             & \multicolumn{3}{c}{$n = 1000$}                                           \\
              & \multicolumn{3}{c}{$n$}                                                  & \multicolumn{3}{c}{$p$}                                                  \\ 
\cmidrule(lr){2-4} \cmidrule(lr){5-7} 
Method        & 500                    & 1000                   & 2000                   & 25                     & 50                     & 100                    \\ 
\hline
& \multicolumn{6}{c}{Spearman correlation in factor score estimation} \\
\cmidrule(lr){2-7}
scFM $(k=4)$  & 0.973 (0.000)          & 0.976 (0.000)         & 0.978 (0.000)          & 0.966 (0.000)          & 0.976 (0.000)          & 0.989 (0.000)          \\
scFM $(k=8)$  & 0.962 (0.001)          & 0.964 (0.000)          & 0.964 (0.000)          & 0.952 (0.000)          & 0.964 (0.000)          & 0.983 (0.001)          \\
ZIFA          & 0.933 (0.001)          & 0.934 (0.001)          & 0.935 (0.000)          & 0.893 (0.004)          & 0.934 (0.001)          & 0.975 (0.000)          \\
ZIPFA         & 0.532 (0.005)          & 0.530 (0.005)          & 0.525 (0.003)          & 0.401 (0.006)          & 0.530 (0.005)          & 0.639 (0.003)          \\
CopulaFM      & 0.916 (0.001)          & 0.889 (0.001)          & 0.851 (0.001)          & 0.929 (0.000)          & 0.889 (0.001)          & N/A                    \\ 
\hline 
& \multicolumn{6}{c}{Spearman correlation in factor loading estimation} \\
\cmidrule(lr){2-7}
scFM $(k=4)$  & 0.991 (0.000)          & 0.994 (0.000)          & 0.995 (0.000)          & 0.995 (0.000)          & 0.994 (0.000)          & 0.995 (0.000)          \\
scFM $(k=8)$  & 0.992 (0.000)         & 0.995 (0.000)          & 0.997 (0.000)          & 0.996 (0.000)          & 0.995 (0.000)          & 0.996 (0.000)          \\
ZIFA          & 0.591 (0.001)          & 0.673 (0.001)          & 0.595 (0.001)          & 0.664 (0.003)          & 0.673 (0.001)          & 0.665 (0.001)          \\
ZIPFA         & 0.156 (0.006)          & 0.153 (0.007)          & 0.143 (0.005)          & 0.139 (0.006)          & 0.153 (0.007)          & 0.440 (0.008)          \\
CopulaFM      & 0.916 (0.001)          & 0.891 (0.001)          & 0.882 (0.001)          & 0.960 (0.000)          & 0.891 (0.001)          & N/A                    \\
\hline
\end{tabular}%
}
\end{center}
\end{table}

The resulting correlation values for the factor scores and loadings are summarized in Tables \ref{tab:scores}. Results for CopulaFM with $p = 100$ are omitted due to numerical issues encountered with its R package implementation. 
The proposed scFM consistently outperforms its competitors in estimating both factor scores and factor loadings, demonstrating its advantages on skewed count data with excessive zero and near-zero values. Notably, even when using a value of $k_{\text{max}}$ larger than the ground truth (i.e., $k_{\text{max}}=8$), the estimation of factor scores by scFM did not deteriorate much and is still the second best across all scenarios. 
Similarly, in the estimation of factor loadings, our scFM outperforms the competing approaches across every scenario for both $k_{\text{max}} \in \{4, 8\}$.  This highlight the efficacy of our column-wise DL prior in adapting to the unknown number of latent factors and the robustness of our method to different $k_{\text{max}}$ selections.
Remarkably, the approaches that utilize the copula model (i.e., scFM and CopulaFM) exhibit superior performance in estimating factor loadings compared to their parametric alternatives. This demonstrates the advantage of employing the copula model framework for factor analysis on datasets characterized by the inflation of low counts and high skewness. 
However, despite the flexibility of CopulaFM as a copula-based factor model, its performance falls short of ours, with its effectiveness diminishing as data and feature sizes grow. CopulaFM  also encounters numerical issues, such as the failure of the Cholesky decomposition, in the scenario with the largest number of genes ($p=100$). We conjecture that larger data and feature sizes exacerbates the multi-modality issue in its posterior distribution, deteriorating the mixing of the MCMC implementation for CopulaFM and resulting in inferior results. This multi-modality issue might also navigate the MCMC algorithm into regions prone to numerical issues.
To summarize, scFM demonstrates highly effective performance in estimating both factor scores and loadings for datasets that exhibit the characteristics of scRNA-seq data -- an abundance of low counts and high skewness.

\section{Application to scRNA-seq data}
\label{sec:reald}

\subsection{Data and aim}
We apply the proposed scFM to our motivating scRNA-seq dataset, which consists of 7,247 cells from the lymphoblastoid cell line (LCL) GM12878, for which 18,170 genes were sequenced. The data can be obtained from the 10x Genomics Datasets website (\url{https://www.10xgenomics.com}).
The goal of this analysis is to characterize underlying relationships among the sequenced genes and to obtain low-dimensional factor scores summarizing the variability in the observed scRNA-seq data to aid detection of previously uncharacterized cell subtypes in LCL.
We filter out low-quality cells such as empty droplets, cell doublets and multiplets using the R package Seurat \citep{Hao2023-ak}, resulting in $n = 5,135$ cells.  The detailed data preprocessing procedure is provided in Appendix A. 
We focus on $p = 100$ genes that show the largest variability across cells, which contain $\sim 58\%$ zeros and $\sim 15\%$ ones across those genes.

\subsection{Factor analysis}\label{sec:reald-fa}

For the proposed scFM, we use $m = 1$ to explicitly account for the notable inflation of zero and one counts observed in the motivating scRNA-seq dataset (refer to Figure \ref{fig:genes_selected}). 
We set the maximum number of latent factors at $k_{\text{max}}=8$ to achieve a balance between model interpretability and avoiding model misspecification, as fewer latent factors tend to provide clearer insights for model interpretation while too few risks excluding significant factors. Our analysis supports this decision, demonstrating that the number of significant latent factors is indeed fewer than $k_{\text{max}}=8$.
We use the same hyperparameter values as in Section \ref{sec:sim} and run the proposed MCMC in Section \ref{sec:infer} for 10,000 iterations with 5,000 burn-in. To assess the goodness-of-fit of the proposed scFM on this dataset, we perform posterior predictive checks, which compare the posterior predictive distribution with the observed data. Our proposed model shows no significant lack of fit; see Appendix A for details.

We utilize the approach outlined in Section \ref{sec:estim_k*} to determine the number of significant factors $\hat{k}$ and to discern their identity. For this dataset, scFM determines that there are two significant factors (i.e. $\hat{k} = 2$), and factors corresponding to the two largest columns (in $\ell_2$-norm) of the posterior mean of the factor loadings matrix $\Lambdab$  are identified as the significant factors (termed factors 1 and 2). 

\begin{figure}
\centering
\begin{subfigure}{\textwidth}
\includegraphics[width=0.9\linewidth]{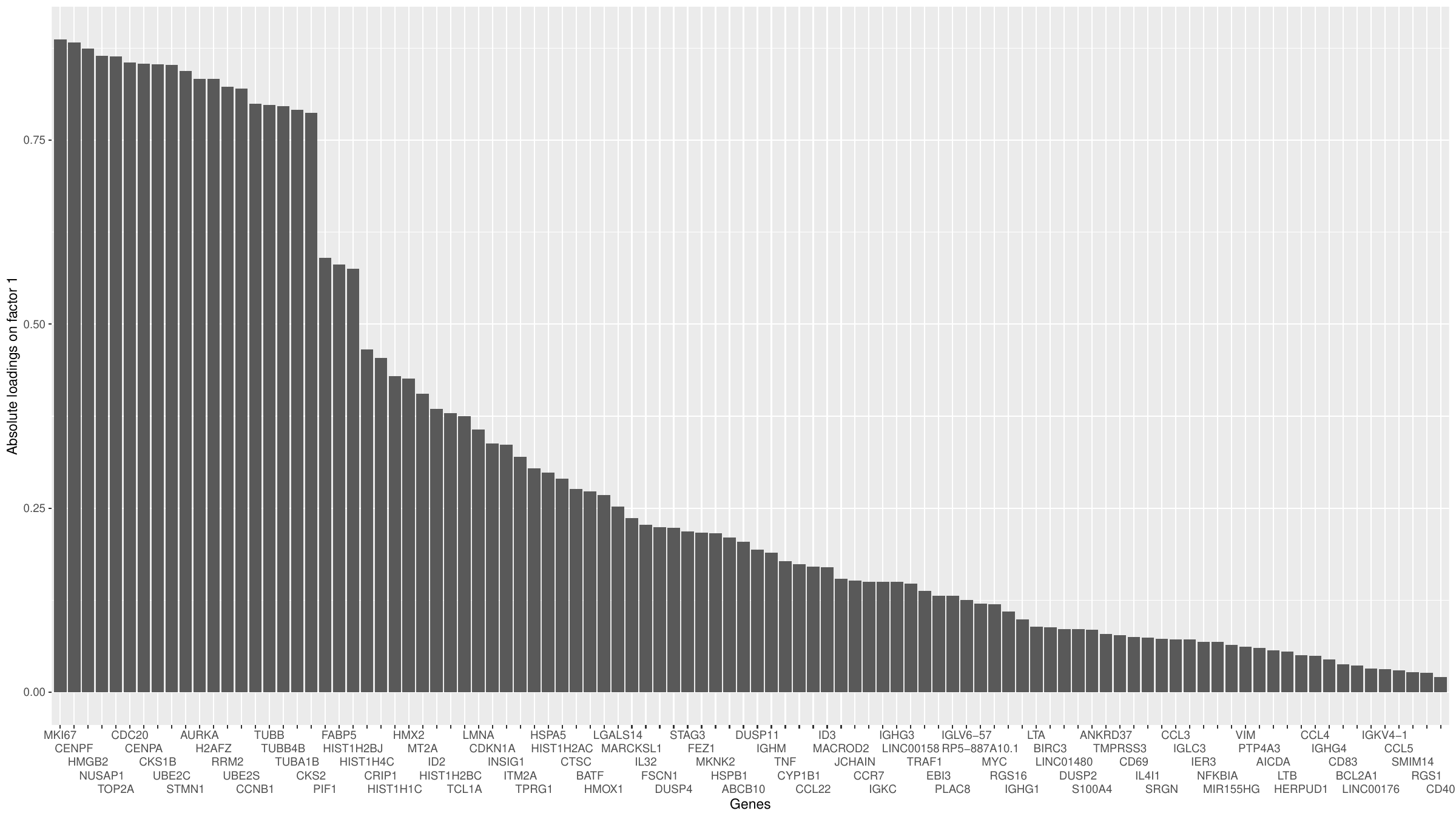}
\caption{Absolute loadings on factor 1}
\end{subfigure}
\begin{subfigure}{\textwidth}
\includegraphics[width=0.9\linewidth]{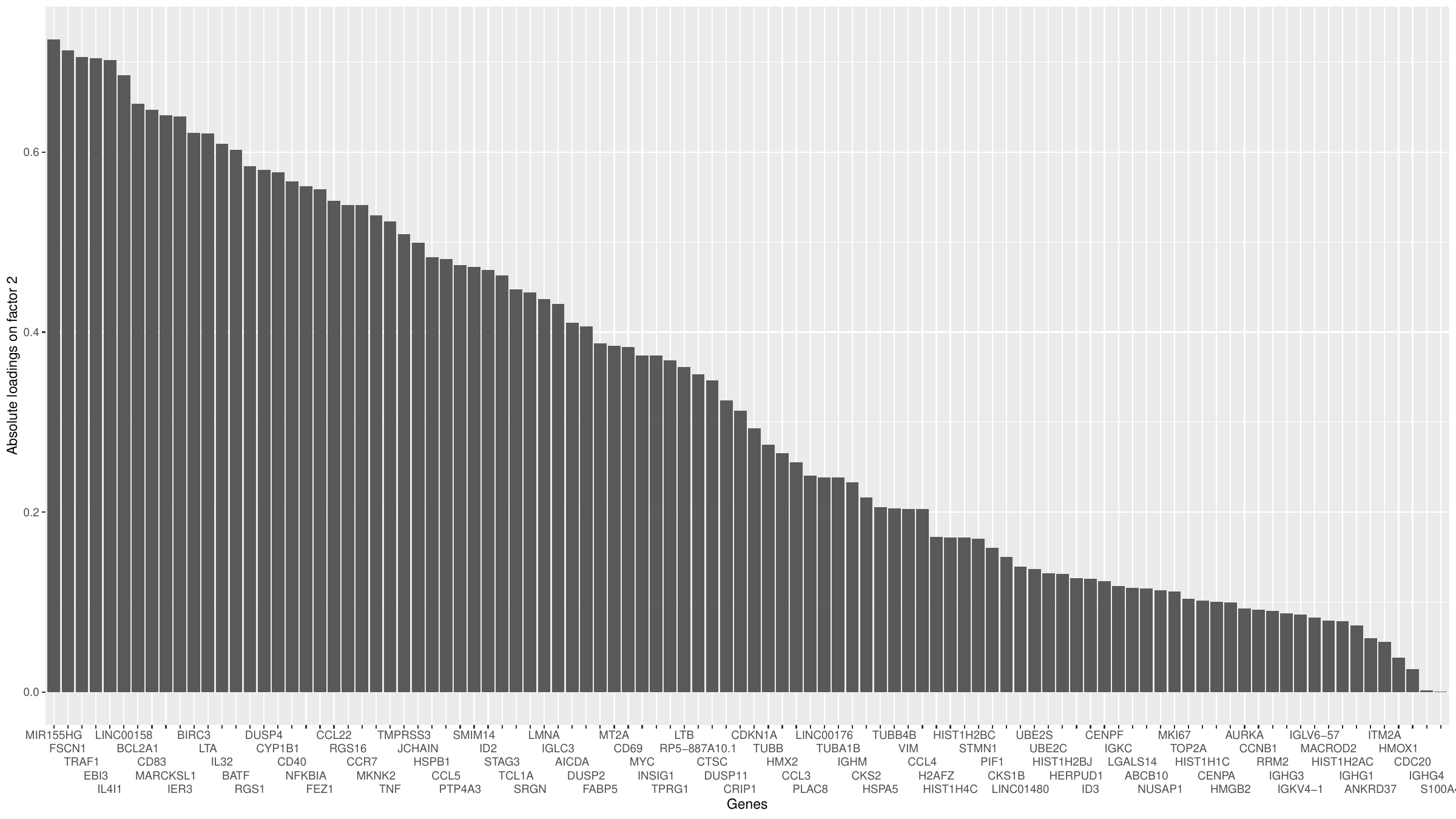}
\caption{Absolute loadings on factor 2}
\end{subfigure}
\caption{\label{fig:loadings} Absolute values of the estimated factor loadings on the significant factors.}
\end{figure}

Figure \ref{fig:loadings} illustrates the absolute loadings of all genes on these two significant factors, providing insight into a potential biological interpretation for each estimated factor. For factor 1, genes with the 10 largest absolute loadings are  \texttt{MKI67}, \texttt{CENPF}, \texttt{HMGB2}, \texttt{NUSAP1}, \texttt{TOP2A}, \texttt{CDC20}, \texttt{CENPA}, \texttt{CKS1B}, \texttt{UBE2C}, and \texttt{STMN1}. Notably, several of these genes are closely linked to the cell cycle, a fundamental process where cells go through a series of events leading to cell division. Specifically, \texttt{MKI67}, \texttt{CENPF}, \texttt{HMGB2}, and \texttt{TOP2A} have been known as cell cycle markers in single-cell studies of human lymphoblastoid cell lines \citep{Sawada2020-tr,SoRelle2022-fh}. Therefore, we call factor 1 the \emph{cell cycle factor}.

For factor 2, genes with the ten most substantial loadings are \texttt{MIR155HG}, \texttt{FSCN1}, \texttt{TRAF1}, \texttt{EBI3}, \texttt{IL4I1}, \texttt{LINC00158}, \texttt{BCL2A1}, \texttt{CD83}, \texttt{MARCKSL1}, and \texttt{IER3}, some of which are known to be related to immune functions. \texttt{FSCN1} and \texttt{CD83} are markers for dendritic cells, specialized immune cells crucial for presenting antigens to other immune cells and initiating immune responses \citep{Zimmer2012-dg,Lechmann2002-lg}. Moreover, \texttt{MIR155HG} and \texttt{IL4I1} are frequently over-expressed in immune cells including B-cells, and play important roles in immune infiltration \citep{Peng2019-tu,Bod2018-kq}. Therefore, we call factor 2 the \emph{immune cell factor}.

\subsection{Cell clustering}
To further interpret the two major latent factors identified by the proposed scFM, we use them to identify distinct cellular subtypes or states in LCL. The scores of the two selected factors serve as a low-dimensional representation of the high-dimensional gene expression measurements in scRNA-seq data. Therefore, by clustering the data in the low-dimensional factor space, we can identify groups of cells exhibiting similar expression behavior. For simplicity, we use the k-means algorithm to cluster the latent factor scores, where the number of clusters is determined by the elbow method. 

\begin{figure}
\centering
\includegraphics[width=0.9\linewidth]{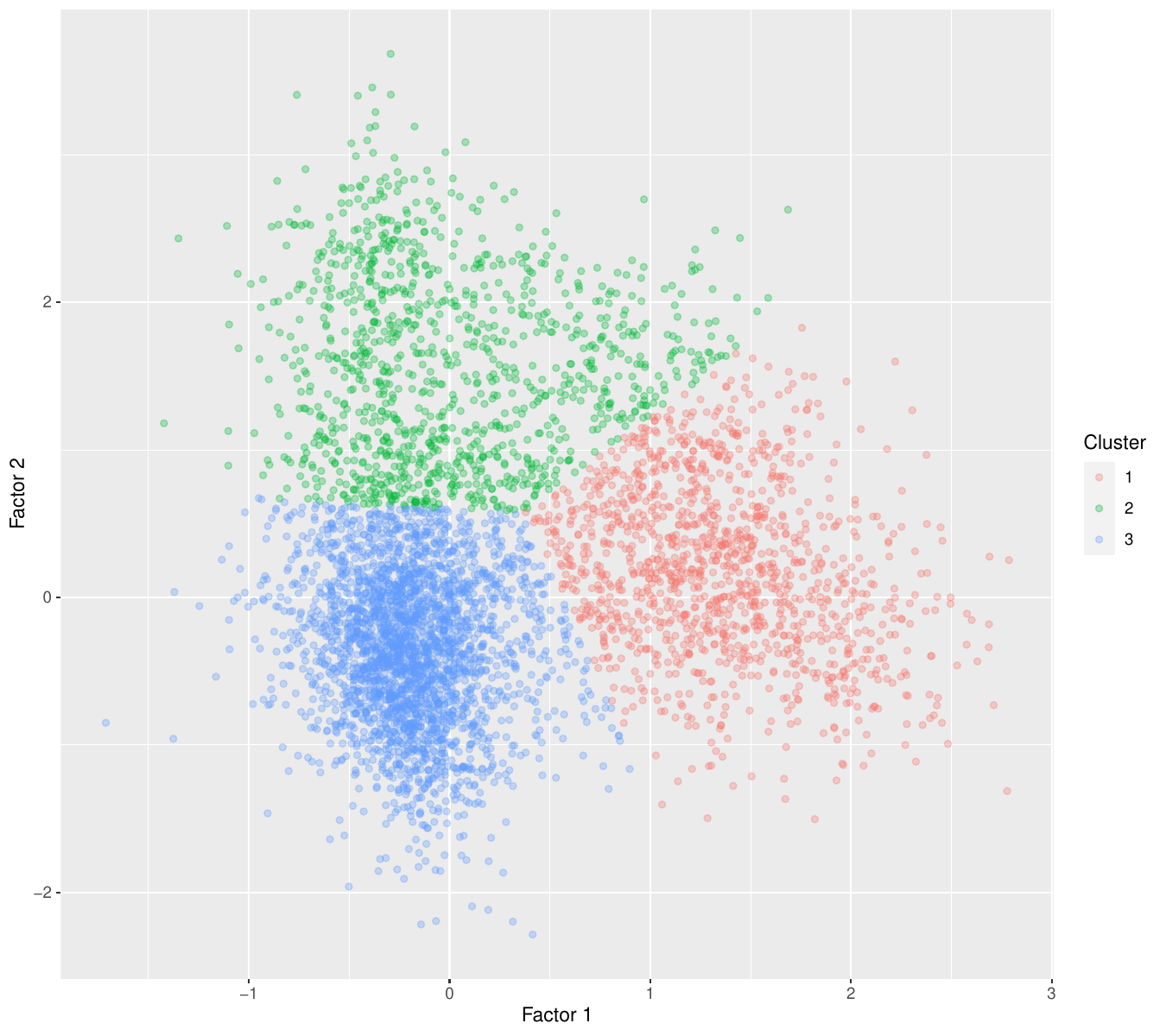}
\caption{\label{fig:clustering} Clusters of cells based on the k-means clustering of scores of factors 1 and 2, for which the elbow method reveals the presence of three clusters. For better visualization, one outlier is excluded. 
}
\end{figure}

Our analysis reveals three clusters, which are depicted in Figure \ref{fig:clustering}. For cluster 1, we observe a notable elevation in scores for factor 1, while scores for factor 2 are close to zero. Cluster 2, in contrast, exhibits elevated scores for factor 2 but with diminished scores for factor 1. Cluster 3 displays factor scores around zero for both factors. 
Combining this observation with the biological interpretation of the estimated factors in Section \ref{sec:reald-fa} suggests that cluster 1 likely comprises cycling cells, cluster 2 likely consists of immune cells, and the remaining cells are grouped into cluster 3. We further investigate gene expression measurements for relevant genes discussed in Section \ref{sec:reald-fa} to support this interpretation. Table \ref{tab:avg_cell_type} reports average gene expression measurements of \texttt{MKI67}, \texttt{CENPF}, \texttt{HMGB2}, \texttt{TOP2A}, \texttt{FSCN1}, \texttt{CD83}, \texttt{MIR155HG}, and \texttt{IL4I1} for the entire dataset and for each cluster. Recall that the first four of these genes are associated with factor 1, while the latter four are associated with factor 2. For cluster 1, \texttt{MKI67}, \texttt{CENPF}, \texttt{HMGB2}, and \texttt{TOP2A} are over-expressed.  Given that \texttt{MKI67}, \texttt{CENPF}, \texttt{HMGB2}, and \texttt{TOP2A} serve as markers for cycling cells, it can be inferred that cells in cluster 1 are currently undergoing cell cycling. On the other hand,  cluster 2 exhibits high expression of the latter four genes \texttt{FSCN1}, \texttt{CD83}, \texttt{MIR155HG}, and \texttt{IL4I1}, which is known to be strongly correlated with immune cells and immune cell function. Consequently, it is plausible to conclude that cluster 2 represents immune cells.

\begin{table}
\caption{Average gene expression levels of \texttt{MKI67}, \texttt{CENPF}, \texttt{TOP2A}, \texttt{FSCN1}, \texttt{CD83}, \texttt{MIR155HG}, \texttt{IL4I1} for the entire dataset, as well as for cluster 1, cluster 2, and cluster 3, respectively.}
\label{tab:avg_cell_type}
\begin{center}
\begin{tabular}{lcccccccc}
\hline
               & \texttt{MKI67} & \texttt{CENPF} & \texttt{HMGB2} & \texttt{TOP2A} & \texttt{FSCN1} & \texttt{CD83} & \texttt{MIR155HG} & \texttt{IL4I1} \\
\hline
Entire dataset & 1.938          & 0.942          & 7.622          & 0.777          & 1.536          & 0.789         & 3.238             & 1.038 \\
Cluster 1      & 6.493          & 3.241          & 24.911         & 2.716          & 1.022          & 0.422         & 1.864             & 0.628 \\
Cluster 2      & 1.147          & 0.469          & 4.390          & 0.366          & 5.585          & 2.786         & 11.276            & 3.653 \\
Cluster 3      & 0.140          & 0.065          & 0.892          & 0.042          & 0.127          & 0.147         & 0.603             & 0.164 \\
\hline
\end{tabular}
\end{center}
\end{table}


\section{Discussion}
\label{sec:disc}

In this work, we have developed a segmented Gaussian copula factor model with application to single-cell sequencing data, which have an excessive amount of low counts due to limited sequencing depth. Our approach adopts a Bayesian framework that utilizes a column-wise Dirichlet-Laplace prior for the factor loadings. This framework accommodates the uncertainty related to the number of latent factors, enabling automatic selection of the most significant ones. Additionally, our formulation resolves the identifiability issues inherent in factor models, thereby enhancing model interpretability. Simulation studies on synthetic datasets that mimic real scRNA-seq data show that the proposed approach significantly outperforms existing methods. In an application to scRNA-seq data, our method finds two biologically meaningful latent factors and identifies distinct cellular subpopulations with differing expression behaviors.

While our primary focus lies in the analysis of low-count inflated data, specifically single-cell sequencing data, our proposed model can be directly extended to mixed type data. An example of such data can be found in projects like the Human Cell Atlas Project, which contains a combination of binary (e.g., mutation) and low-count inflated data (e.g., scRNA-seq). We expect that extending our model to address such mixed data would be feasible by leveraging the Gaussian copula model for mixed data as proposed by \cite{Fan2016-sc} along with the proposed segmented Gaussian copula model.


\section*{Acknowledgements}
Ni's research is partially supported by CPRIT RP230204, NIH 1R01GM148974-01, and NSF DMS-2112943. Gaynanova's research is partially supported by NSF DMS-2044823.
\vspace*{-8pt}

\bibliographystyle{chicago}
\bibliography{references}

\appendix
\counterwithin{figure}{section}
\renewcommand{\thefigure}{A.\arabic{figure}}

\section{Additional details for application to scRNA-seq data}
\subsection{Data pre-processing}

\begin{figure}[t]
\centering
\includegraphics[width=\linewidth]{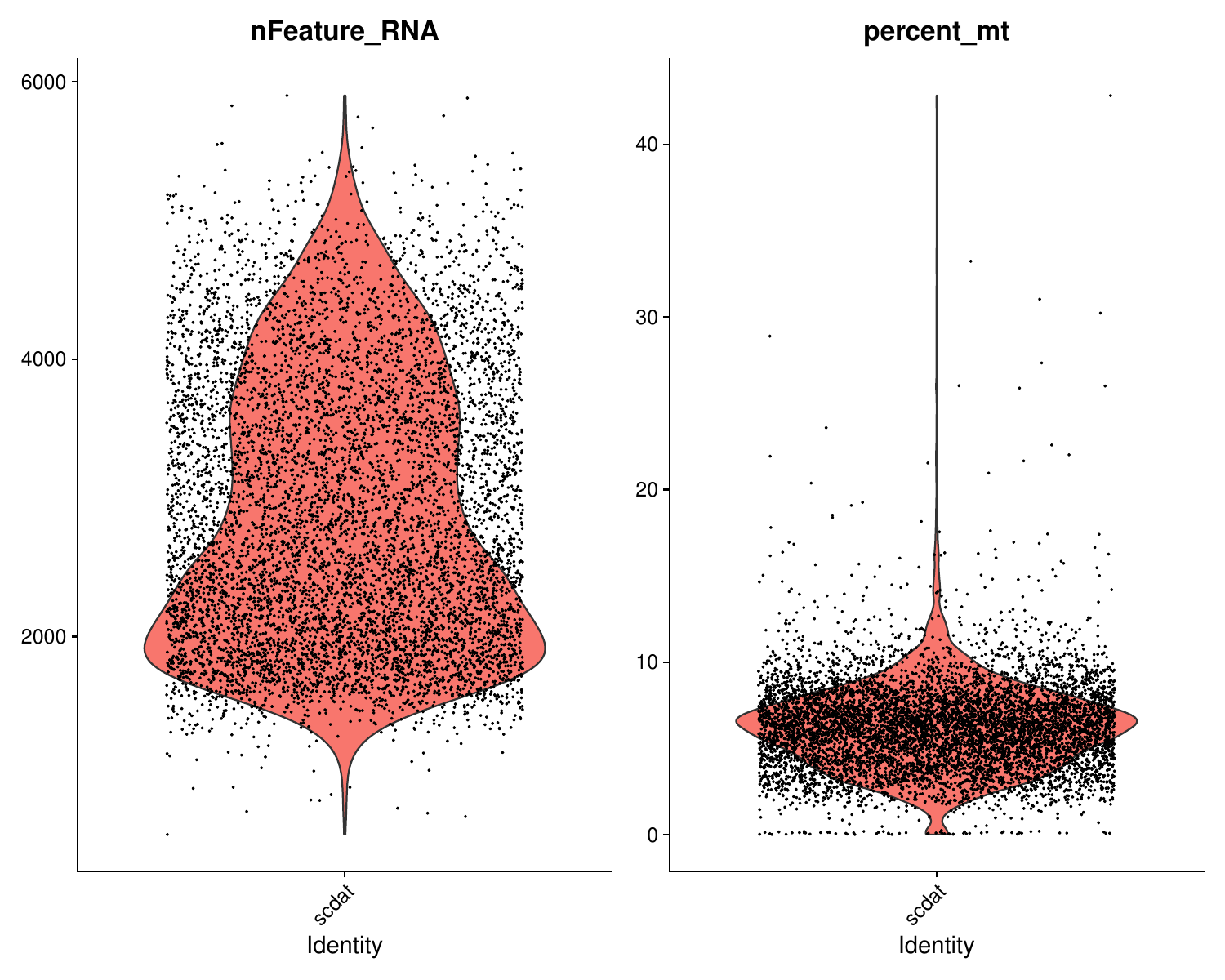}
\caption{\label{webfig:qc_metrics} Violin plots displaying the count of unique genes identified in individual cells (nFeature\_RNA) and the percentage of sequencing reads mapping to the mitochondrial genome (percent\_mt) in individual cells.}
\end{figure}

The scRNA-seq dataset in Section 5 is pre-processed using the \texttt{R} package \texttt{Seurat} \citep{Hao2023-ak} for our factor analysis. After eliminating genes with over $90\%$ zeros, we further refine the dataset by filtering out low-quality cells such as empty droplets, cell doublets, and multiplets based on quality control (QC) metrics. We then select $p = 100$ genes showing the largest cell-to-cell variation. Among the commonly used QC metrics for scRNA-seq data, we specifically consider the number of unique genes detected in each cell and the percentage of reads mapping to the mitochondrial genome, which are illustrated in Figure \ref{webfig:qc_metrics}.
Cells of poor quality or empty droplets typically exhibit a very low gene count, while cell doublets or multiplets often display an unusually high gene count. Additionally, low-quality cells frequently demonstrate significant mitochondrial contamination. Consequently, we filter out cells with unique feature counts over $4,200$ or less than $1,800$, as well as cells with mitochondrial counts exceeding $10\%$. Based on the retained $n = 5,133$ cells, we identify $p=100$ genes exhibiting the most substantial cell-to-cell variation in the dataset:

\noindent \texttt{IGHG1}, \texttt{IGHG3}, \texttt{IGKC}, \texttt{CCL22}, \texttt{IGHM}, \texttt{CCL3}, \texttt{IGHG4}, \texttt{CCL4}, \texttt{IGLC3}, \texttt{HIST1H1C}, \texttt{HIST1H2BJ}, \texttt{LTB}, \texttt{LTA}, \texttt{HSPB1}, \texttt{CYP1B1}, \texttt{RGS1}, \texttt{HIST1H4C}, \texttt{LINC00176}, \texttt{CD69}, \texttt{MIR155HG}, \texttt{IGLV6-57}, \texttt{UBE2C}, \texttt{CTSC}, \texttt{FSCN1}, \texttt{MARCKSL1}, \texttt{NFKBIA}, \texttt{CCR7}, \texttt{MT2A}, \texttt{DUSP2}, \texttt{CD83}, \texttt{ABCB10}, \texttt{IL4I1}, \texttt{IER3}, \texttt{JCHAIN}, \texttt{UBE2S}, \texttt{LMNA}, \texttt{DUSP4}, \texttt{IL32}, \texttt{HMGB2}, \texttt{RGS16}, \texttt{HMOX1}, \texttt{BIRC3}, \texttt{TMPRSS3}, \texttt{S100A4}, \texttt{MYC}, \texttt{TNF}, \texttt{CDKN1A}, \texttt{TUBA1B}, \texttt{TCL1A}, \texttt{CRIP1}, \texttt{BCL2A1}, \texttt{TOP2A}, \texttt{EBI3}, \texttt{CDC20},
\texttt{ANKRD37}, \texttt{HMX2}, \texttt{AICDA}, \texttt{ID3}, \texttt{MKI67}, \texttt{LGALS14}, \texttt{TPRG1}, \texttt{STAG3}, \texttt{CCL5}, \texttt{HIST1H2BC}, \texttt{CKS2}, \texttt{RRM2}, \texttt{HIST1H2AC}, \texttt{CCNB1}, \texttt{IGKV4-1}, \texttt{FABP5}, \texttt{CENPF}, \texttt{MACROD2}, \texttt{CENPA}, \texttt{H2AFZ}, \texttt{INSIG1}, \texttt{MKNK2}, \texttt{RP5-887A10.1}, \texttt{SRGN}, \texttt{LINC01480}, \texttt{LINC00158}, \texttt{HSPA5}, \texttt{TUBB4B}, \texttt{PIF1}, \texttt{ITM2A}, \texttt{CD40}, \texttt{TUBB}, \texttt{VIM}, \texttt{PLAC8}, \texttt{DUSP11}, \texttt{STMN1}, \texttt{CKS1B}, \texttt{HERPUD1}, \texttt{TRAF1}, \texttt{PTP4A3}, \texttt{NUSAP1}, \texttt{BATF}, \texttt{ID2}, \texttt{FEZ1}, \texttt{AURKA}, \texttt{SMIM14}.

\subsection{Goodness-of-fit}

We perform posterior predictive checks \citep{Rubin1984-bu} to assess the goodness-of-fit of the proposed scFM on the scRNA-seq dataset in Section 5. To assess model fit, these checks compare observed data with the posterior predictive distribution of hypothetical replicated data that reflects what we expect under our posterior distribution. A good fit is indicated when the observed data aligns with the posterior predictive distribution based on the model.

\begin{figure}[t]
\centering
\includegraphics[width=\linewidth]{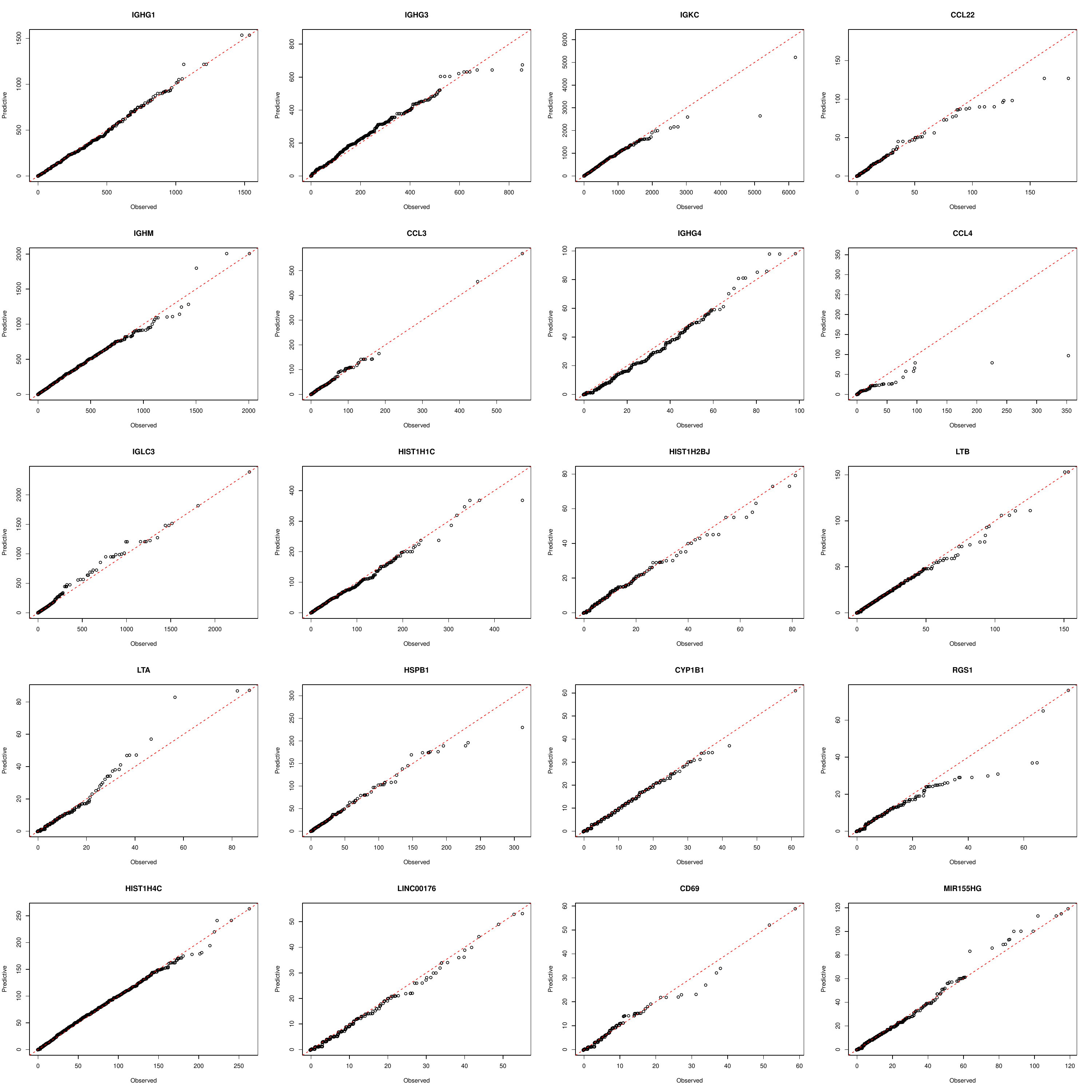}
\caption{\label{webfig:ppc1} The QQ-plots corresponding to genes 1-20, with observed and posterior predictive quantiles represented on the X and Y axes, respectively.}
\end{figure}

\begin{figure}[!t]
\centering
\includegraphics[width=\linewidth]{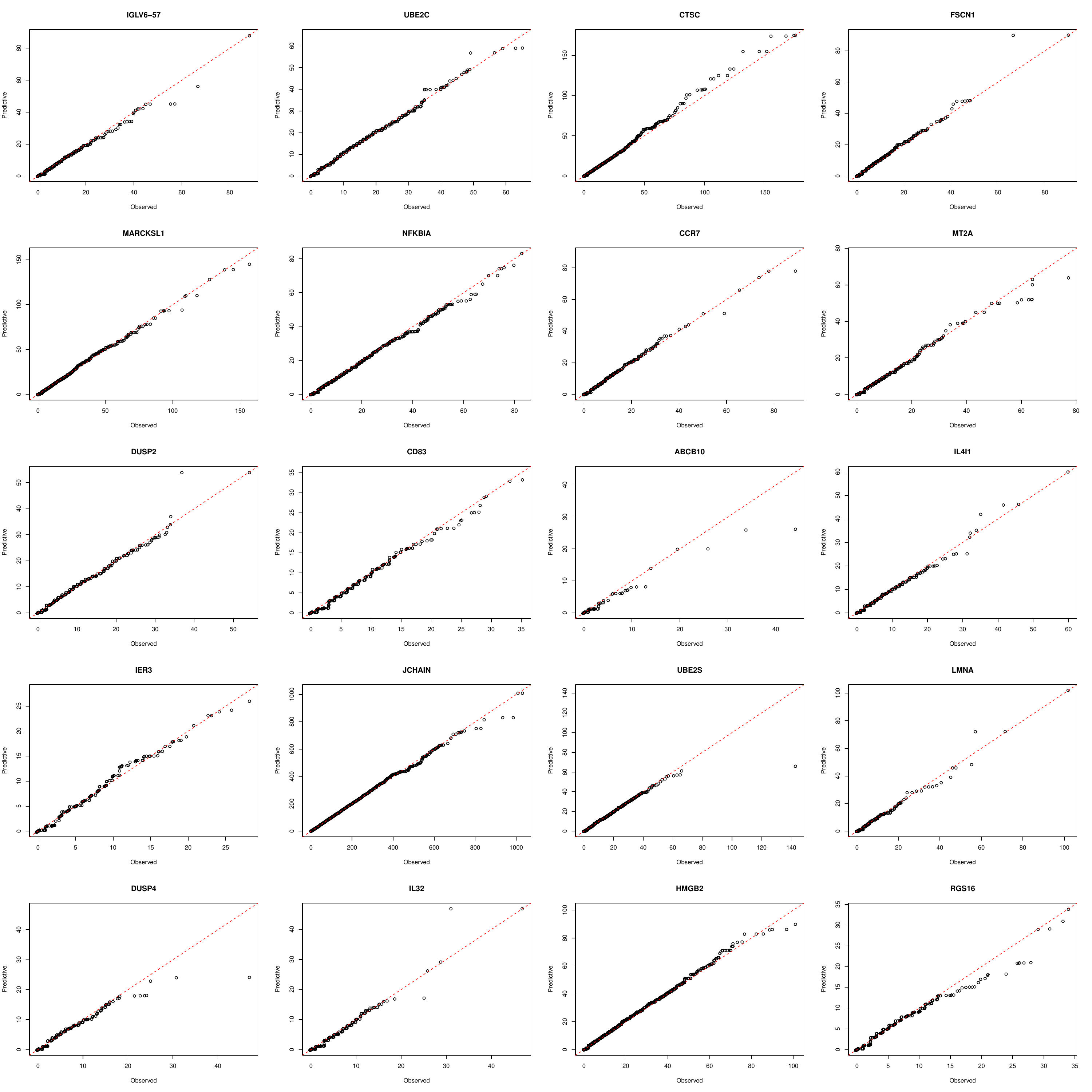}
\caption{\label{webfig:ppc2} The QQ-plots corresponding to genes 21-40, with observed and posterior predictive quantiles represented on the X and Y axes, respectively.}
\end{figure}

\begin{figure}[!t]
\centering
\includegraphics[width=\linewidth]{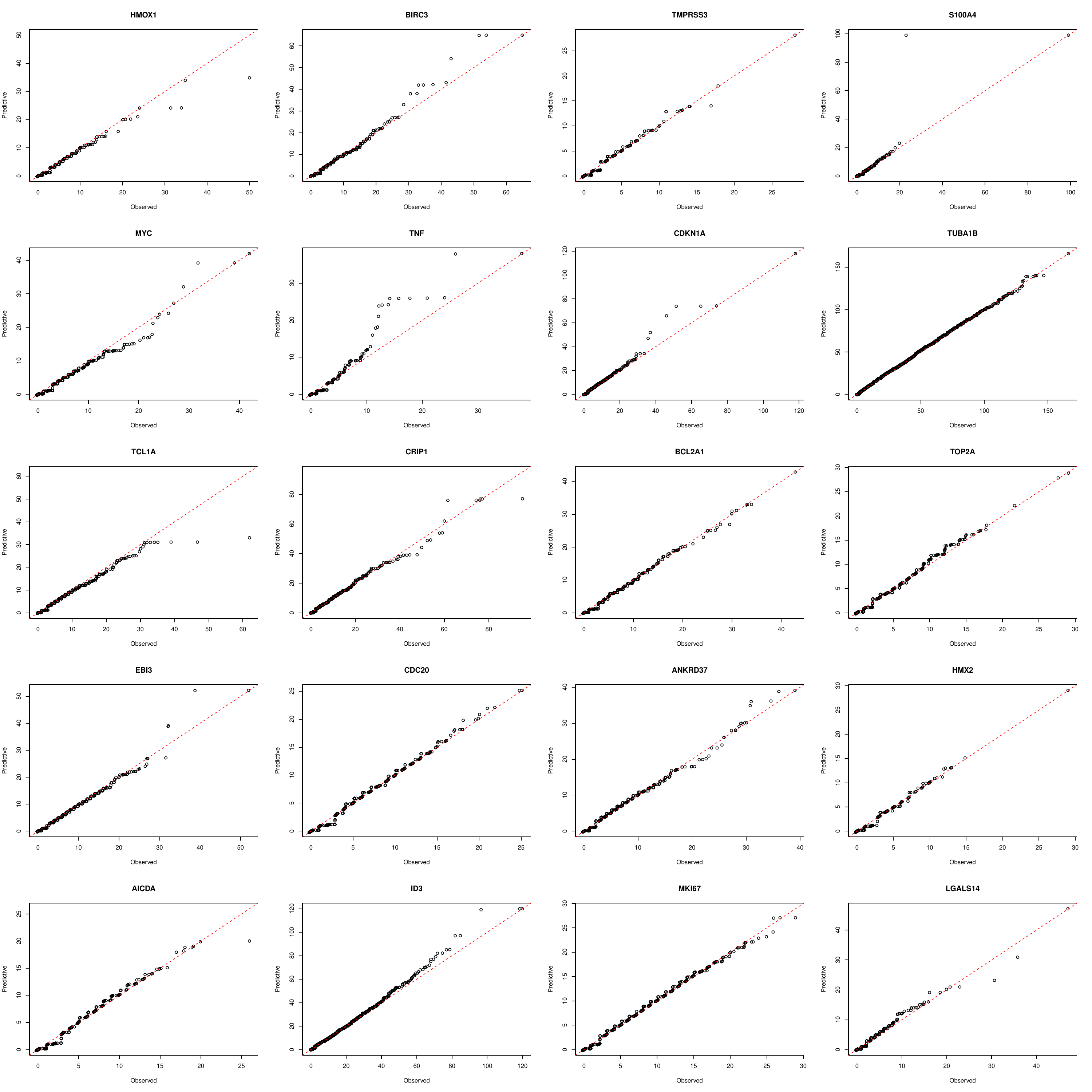}
\caption{\label{webfig:ppc3} The QQ-plots corresponding to genes 41-60, with observed and posterior predictive quantiles represented on the X and Y axes, respectively.}
\end{figure}

\begin{figure}[!t]
\centering
\includegraphics[width=\linewidth]{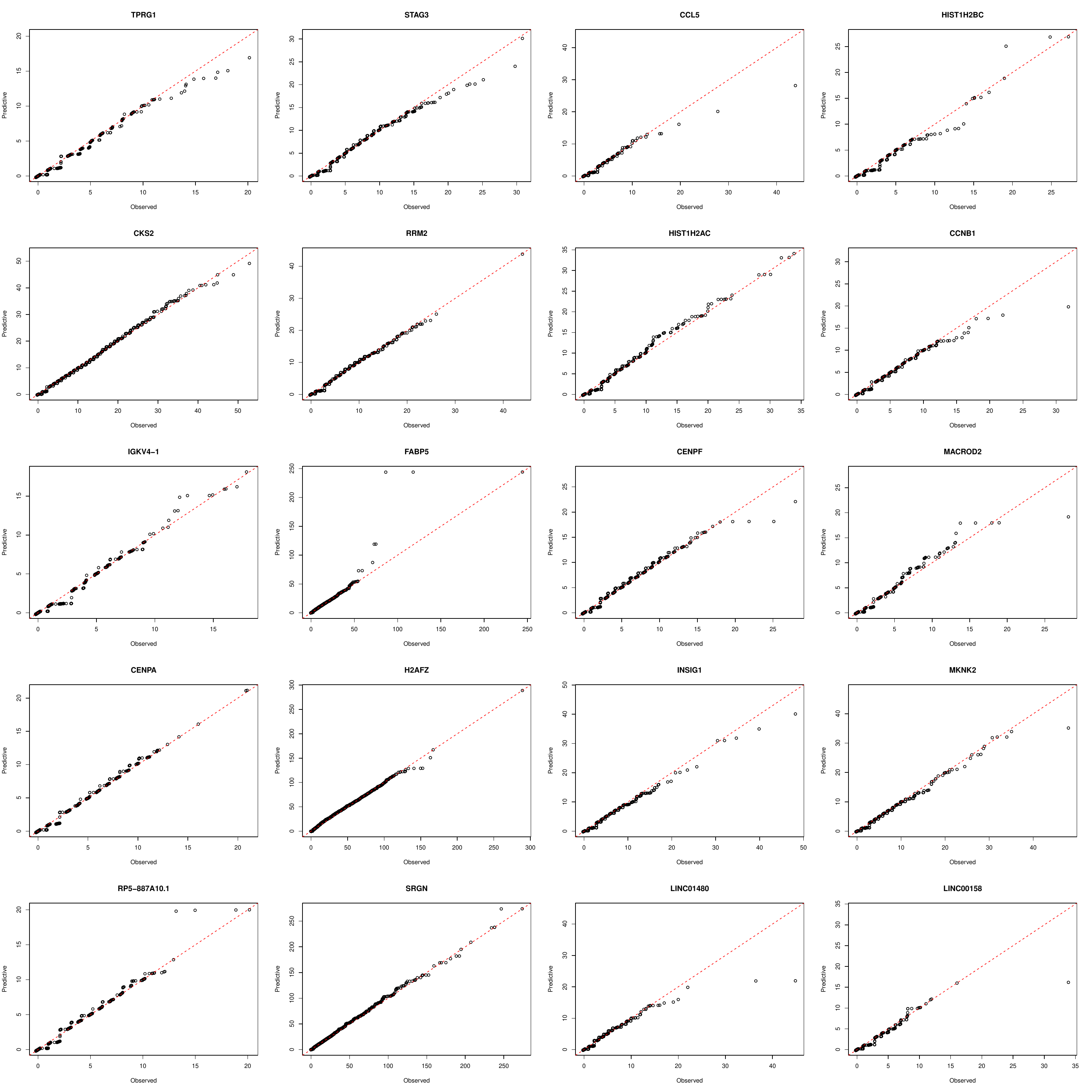}
\caption{\label{webfig:ppc4} The QQ-plots corresponding to genes 61-80, with observed and posterior predictive quantiles represented on the X and Y axes, respectively.}
\end{figure}

\begin{figure}[!t]
\centering
\includegraphics[width=\linewidth]{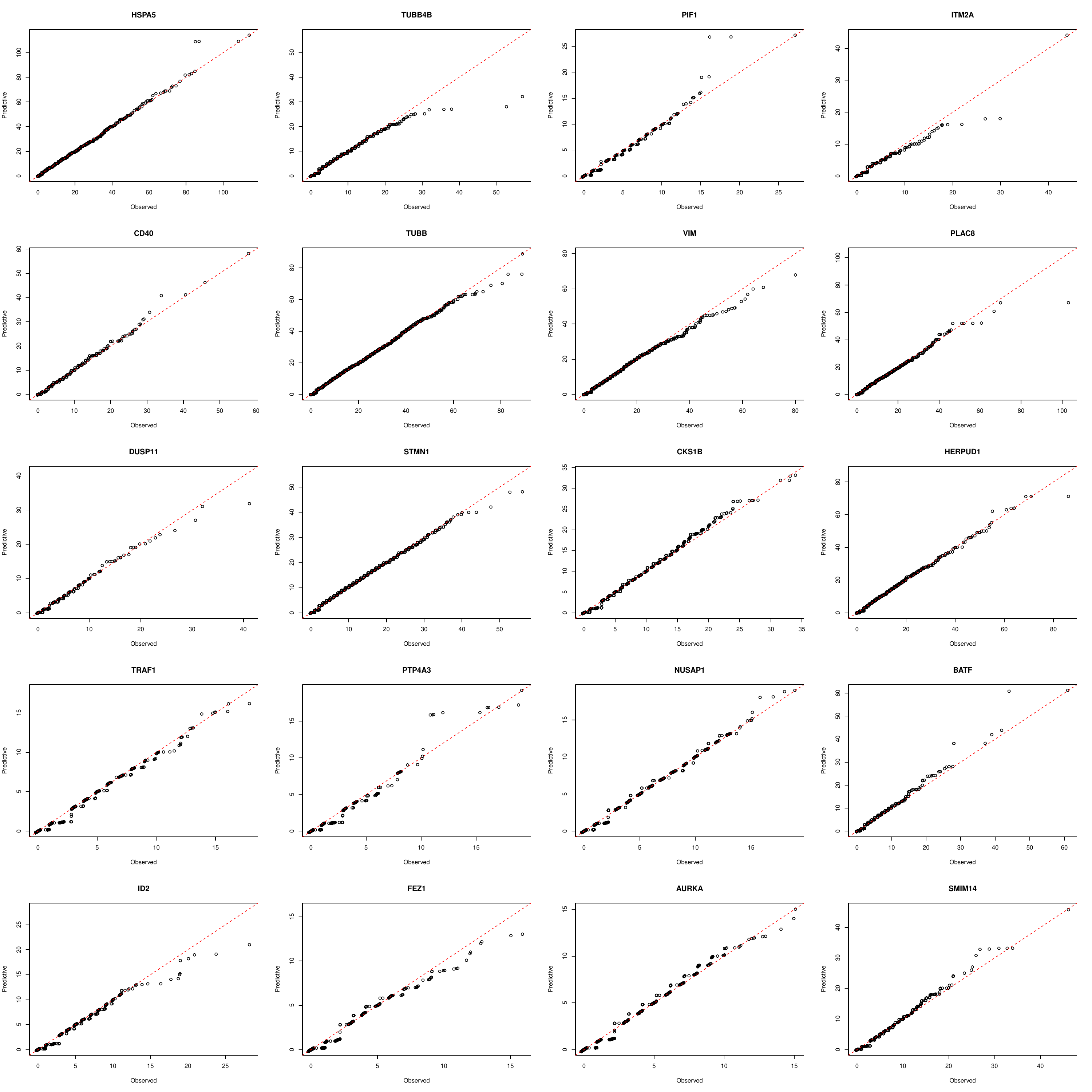}
\caption{\label{webfig:ppc5} The QQ-plots corresponding to genes 81-100, with observed and posterior predictive quantiles represented on the X and Y axes, respectively.}
\end{figure}

Specifically, we generate posterior predictive samples $\bm{x}^{(s)}, s = 1, \ldots, S$ as follow:
\begin{align*}
\zb^{(s)} & \sim \N_p(\bm{0}, \Omegab^{(s)}), \\
x_j^{(s)} & = \sum_{d=0}^1 1(\delta_{jd}^{(s)} < z_j^{(s)} \le \delta_{j,d+1}^{(s)}) d + 1(z_j^{(s)} > \delta_{j2}^{(s)}) \hat{F}_j^{-1} (\Phi (z_j^{(s)})), 
\end{align*}
where $\Omegab^{(s)}$ is the correlation matrix obtained from the covariance matrix $\Lambdab^{(s) \top} \Lambdab^{(s)} + \Sigmab^{(s)}$, which is constructed using posterior samples $\{\Lambdab^{(s)} , \Sigmab^{(s)}\}$ of $(\Lambdab, \Sigmab)$. Moreover, $\hat{F}_j^{-1}$ denotes the pseudo inverse of the empirical cdf $\hat{F}_j$ and $\delta_{jd}^{(s)}$ are posterior samples of the Gaussian level thresholds $\delta_{jd}$.
We compare these posterior predictive samples with the observed data for each gene. In general, the posterior predictive samples show very good fits to the observed data, as illustrated in the Q-Q plots in Figures \ref{webfig:ppc1}-\ref{webfig:ppc5}. For each gene, the majority of data points closely align along  the 45-degree line, with only a few deviations observed at the tail, suggesting excellent fits to the observed data.

\end{document}